# The Computational Complexity of Financial Networks with Credit Default Swaps*

Steffen Schuldenzucker          Sven Seuken          Stefano Battiston

University of Zurich          University of Zurich          University of Zurich

### Abstract

The 2008 financial crisis has been attributed to "excessive complexity" of the financial system due to financial innovation. We employ computational complexity theory to make this notion precise. Specifically, we consider the problem of clearing a financial network after a shock. Prior work has shown that when banks can only enter into simple *debt contracts* with each other, then this problem can be solved in polynomial time. In contrast, if they can also enter into *credit default swaps (CDSs)*, i.e., financial derivative contracts that depend on the default of another bank, a solution may not even exist.

In this work, we show that deciding if a solution exists is NP-complete if CDSs are allowed. This remains true if we relax the problem to $\varepsilon$-approximate solutions, for a constant $\varepsilon$. We further show that, under sufficient conditions where a solution is guaranteed to exist, the approximate search problem is PPAD-complete for constant $\varepsilon$. We then try to isolate the "origin" of the complexity. It turns out that already determining which banks default is hard. Further, we show that the complexity is not driven by the dependence of counterparties on each other, but rather hinges on the presence of so-called *naked* CDSs. If naked CDSs are not present, we receive a simple polynomial-time algorithm. Our results are of practical importance for regulators' stress tests and regulatory policy.

## 1  Introduction

The year 2008 has provided a painful example of how moderate losses in a comparatively small financial market can spread and amplify in the financial system to create the worst economic crisis since the Great Depression. It has since become widely accepted that this was not just the result of financial institutions' individual risk-taking, but a consequence of the overall architecture of the financial system at the time. Haldane (2009), then Executive Director of Financial Stability at the Bank of England, described the crisis as a manifestation of "the behaviour under stress of a complex, adaptive network," in which "financial innovation [had] increased further network dimensionality, complexity, and uncertainty." Yellen (2013), then Vice Chair of the Federal Reserve, described regulatory changes after the crisis that are explicitly targeted at "excessive systemic risk arising from the complexity and interconnectedness that characterize our financial system."

---

*Steffen Schuldenzucker, Sven Seuken: Department of Informatics, University of Zurich, Switzerland, Email: {schuldenzucker,seuken}@ifi.uzh.ch. Stefano Battiston: Department of Banking and Finance, University of Zurich, and Swiss Finance Institute, Switzerland, Email: stefano.battiston@uzh.ch. Some of the ideas presented in this paper were previously described in a one-page abstract that was published in the conference proceedings of EC'16 (Schuldenzucker, Seuken and Battiston, 2016) and in a paper that was published in the conference proceedings of ITCS'17 (Schuldenzucker, Seuken and Battiston, 2017).

The financial system has undergone rapid change since the 1990s. *Financial derivatives* (i.e., financial contracts where an obligation to pay depends on some other event or market variable) today allow to trade and reallocate individual components of risk. For example, an investment bank may bundle a loan in a foreign currency with a derivative that will pay the difference between the domestic and the foreign exchange rate. A domestic investor could then buy the bundle without having to worry about a devaluation of the foreign currency. Another trader would take on the other side of the derivative and thus the exchange rate risk. Of course, now both the bank and the investor depend on the trader meeting her obligation to pay. If the trader or the investor is another financial institution, the process can continue over any number of stages, with each party buying, rebundling, and reselling risk. A network of obligations arises: a graph where the nodes are financial institutions ("banks" for short) and the edges are financial contracts. We call this the *financial network*.

The above accounts by policymakers attribute the financial crisis to excessive "complexity" of the financial network. The question remains, though, how exactly we should understand the term "complexity" here. In particular, while a financial network could arise in a variety of ways,[1] most people share an intuition that financial networks with derivatives are "more complex" than ones without. In this paper, we will show that this informal notion materializes in the form of *computational* complexity of a concrete problem that regulators need to solve.

More in detail, we study the *clearing problem*. We are given a financial network consisting of banks and contracts between banks. Each contract defines an obligation to pay a certain amount of money under certain conditions. We assume that some of the banks experienced a shock on their assets, which may render them unable to meet their obligations towards other banks and force them into bankruptcy (or *default*). Defaults may trigger defaults of other banks downstream. For each bank, we are now looking for its *recovery rate*, i.e, the percentage of its liabilities the bank can pay to its creditors. These payments are made from its own (external) assets and the money it receives from other banks. Recovery rates must be in accordance with standard bankruptcy regulations, which imply a constraint reminiscent of a flow identity: defaulting banks must pay out all their assets to creditors and must do so in proportion to the respective obligation. Banks may further incur *default costs* and lose a percentage of their assets upon defaulting. The clearing problem is non-trivial because the contractual relationships can form cycles in the network.[2]

---

[1]Short-term loans between banks and *securization*, i.e., the pooling and re-selling of debt, are two other ways how a financial network can form. The products resulting from securization (most prominently *collateralized debt obligations (CDOs)*) are sometimes called "derivatives," but in contrast to the kinds of derivatives we discuss in this paper, they are defined using a priority structure and do not depend on any market variable except for the debt they are based on. For the purpose of our discussion, we therefore consider these products a form of debt.

The clearing problem serves as a model for how a financial crisis will turn out following the initial shock. Once a solution to the clearing problem has been found, the effect of the initial shock can be judged by metrics like the number of defaulted banks or the total loss of money due to default costs. Researchers have used this approach to study the effect of network structure on systemic risk, such as financial diversification and integration (Elliott, Golub and Jackson, 2014), the interplay of shock size and network structure (Acemoglu, Ozdaglar and Tahbaz-Salehi, 2015), and to determine bounds on the extent of a crisis (Elsinger, Lehar and Summer, 2006; Glasserman and Young, 2015), among many others.[3]

The clearing problem has practical relevance in the context of *stress tests*, where regulators such as the European Central Bank (ECB) evaluate the stability of the financial system under an array of adverse economic scenarios. While today, the official stress tests still operate at a *microprudential* (individual bank) level, efforts are underway to transition to a *macroprudential* point of view, where the financial system is considered as a whole (Constâncio, 2017). These new stress tests need to take network effects into account. For example, the ECB's *STAMP€* framework (Dees, Henry and Martin, 2017) includes a model for the assessment of interbank contagion that is very close to the literature on clearing. The clearing problem is solved many thousands of times in the context of a Monte Carlo simulation. Therefore, these stress tests crucially depend on having access to efficient algorithms for the clearing problem.[4]

Eisenberg and Noe (2001) showed that, if banks can only enter into simple *debt*

---

[2] Perhaps contrary to intuition, cyclic structures are overwhelmingly common in real financial networks. See D'Errico et al. (2018), for example, for an empirical study of the network structure of credit default swap markets, which will be discussed below.

[3] While the exact rules of clearing vary across the literature, they all share as a common feature that payments (or, in some cases, contract valuations) happen simultaneously and some kind of input-output identity needs to hold at each bank. The essential properties of these models with respect to existence and computation of a solution are very similar. In this paper, unless specifically indicated, the term "clearing" refers to the model by Eisenberg and Noe (2001), discussed below, and its extensions.

[4] The size of the financial systems analyzed in stress tests must be expected to lie in the order of 100 banks, making brute-force approaches impractical. For example, Dees, Henry and Martin (2017, Chapter 12) consider 144 banks and the European Banking Authority (2014) stress tests covered 123 banks. Only few of these banks are trivial, i.e., sources or sinks in the network Dees, Henry and Martin (2017, Chart 12.1).

*contracts* with each other, then the clearing problem always has a solution and it can be computed in polynomial time. A debt contract is any contract where the obligation to pay is a fixed number. This could be just a loan from one bank to another, but it can also serve as a model for, e.g., a derivative when the obligation to pay only depends on variables that are external to the financial system and that can assumed to be fixed for the purpose of clearing. Rogers and Veraart (2013) extended the result to default costs. These clearing models have seen widespread adoption in research and stress testing, such as in the above-mentioned *STAMP€* framework. We argue, however, that it is necessary to consider extensions of the model, where the obligation to pay may not be fixed.

Specifically, we study financial networks that contain *credit default swaps (CDSs)* in addition to debt. A CDS is a financial derivative where the obligation to pay depends on the default of a third party, the *reference entity*. Market participants use CDSs to insure themselves against a default of the reference entity or to place a speculative bet on this event. CDSs have played a major role in the default of Lehman Brothers and the bailout of AIG during the 2008 crisis (Fender and Gyntelberg, 2008). Both firms were among the most important institutions in this market, both as counterparties and as reference entities (Fitch Ratings, 2007). It has hence become conventional wisdom that they were counterparties in significant amounts of CDSs where the respective other bank was the reference entity. Such CDSs on other banks cannot accurately be modeled as debt contracts because they depend on an event that is fundamentally endogenous to the financial system. At the same time, future stress tests cannot afford to neglect the dependencies implied by a Lehman–AIG type situation. That is why it is necessary to consider an extension of the existing debt-only to CDSs where reference entities can be other financial institutions.[5]

We consider such a model that extends the Rogers and Veraart (2013) clearing model to CDSs in a straightforward way. In our own recent work (Schuldenzucker, Seuken and Battiston, 2019), we have studied existence of a solution in this model. We have found that there are financial systems where the clearing problem has no solution. At the same time, the clearing algorithms for debt-only networks do not extend to CDSs even in cases where a solution is known to exist. This immediately raises two questions regarding the computational aspects of the clearing problem with CDSs:

1. Given a financial network, can we efficiently determine whether a solution to the clearing problem exists?

2. Given a financial network in which a solution is known to exist, can we efficiently

---

[5] The market for CDSs on financial firms alone currently has a size of about USD 900 billion. In the years following the 2008 crisis, this number was as high as USD 5 trillion. See Bank for International Settlements (2018, Section Single-name instruments, Subsection Financial firms) and the graph linked there.

compute it?

In this paper, we answer both questions in the negative. Towards the first question, we show that it is NP-complete to distinguish networks that have an (exact) solution from those that have no $\varepsilon$-approximate solution, for a natural approximate solution concept and sufficiently small constant $\varepsilon$. In particular, deciding existence of an exact solution or an $\varepsilon$-approximate solution is NP-hard (Section 3).

Towards the second question regarding the computation of a solution, we restrict our attention to the special case where banks do not incur default costs. Here, it is known that a solution always exists (Schuldenzucker, Seuken and Battiston, 2019), but the only known proof of this statement is non-constructive via a fixed-point theorem and so the question regarding computation has remained open so far. As exact solutions can be irrational, we need to consider an approximation problem. We show that the total search problem of finding an $\varepsilon$-approximate solution in a financial system without default costs is PPAD-complete if $\varepsilon$ is a sufficiently small constant. Thus, no polynomial-time approximation scheme (PTAS) exists unless P=PPAD (Section 4).

At this point, we have shown that financial networks with CDSs are indeed "more complex" than those without. However, we should also be able to explain where this newfound complexity comes from. This is important to be able to inform future decisions on regulatory policy beyond an overly simple statement like "CDSs are problematic."

In our quest for an "origin" of the computational complexity, we proceed in two steps. In a first step, we ask what aspect of a solution to the clearing problem is hard to compute. We show that hardness in the decision and search problems does not arise exclusively from the need to compute precise numerical values for the recovery rates. Instead, it is already NP-hard to decide if some given bank will default in some $\varepsilon$-solution (an appropriate distinction variant is NP-*complete*) and in the case without default costs, it is already PPAD-complete to find a set of banks that will default in some $\varepsilon$-solution (Section 5).

In a second step, we study restrictions on the network structure to discern what economic aspects of financial networks the computational complexity might originate from. It follows from our reductions that the problems are still hard in a model where *counterparty risk* (i.e., the dependence of a bank on its debtors) is neglected. Thus, we can say that the complexity originates from *fundamental risk* (i.e., the dependence of CDS counterparties on the reference entity). Finally, we obtain an upper bound on complexity. We show that hardness hinges on the presence of *naked CDSs*, i.e., CDSs that are held without also holding a corresponding debt contract.[6] If naked CDSs are not allowed, a solution always exists and we show that a simple iterative algorithm first presented in Schuldenzucker, Seuken and Battiston (2019) constitutes a fully polynomial-time approximation scheme (FPTAS; Section 6). These insights will allow us to frame a rather complete picture regarding the "origin of the complexity" and they have various implications for regulatory policy (Section 7).

---

[6] Naked CDSs are a common phenomenon in practice. While we are not aware of any empirical studies that quantify the share of CDSs that are naked, there seems to be a broad consensus that they form the majority of CDS positions. Kiff et al. (2009) noted that the (gross) notional of CDSs

Attempts to capture the "complexity" of the financial network have, of course, been made before. We differentiate between informal complexity due to i) the structure of interconnections and ii) the nature of the contracts themselves. Complexity due to the network structure has previously been approached using various measures from graph theory, such as the length of a path between ultimate borrowers and lenders (Shin, 2010), average degree (Gai, Haldane and Kapadia, 2011), network concentration (Arinaminpathy, Kapadia and May, 2012), network entropy (Battiston et al., 2016), or spectral measures (Bardoscia et al., 2017). As these measures require ordinary graphs as their inputs, where edges cannot contain more information than weights, they need to abstract over details of the contracts, such as the dependence of a financial derivative on its underlying market variable. Sensitivity results (Hemenway and Khanna, 2016; Liu and Staum, 2010; Feinstein et al., 2017) are another way to capture "complexity due to interconnectedness" and are also related to computational complexity. These results have so far only been obtained for networks of debt or cross-holdership. Basel III regulations measure a bank's "interconnectedness" by the size of its intra-financial assets, while its "complexity" (of individual contracts) is measured by its amount of OTC[7] derivatives, among others (Basel Committee on Banking Supervision, 2014).

The second kind of complexity, due to the nature of individual contracts, has begun to receive attention from theoretical computer science. Arora et al. (2011) and Zuckerman (2011) studied the cost of asymmetric information in financial derivatives markets with computationally bounded agents. Braverman and Pasricha (2014) showed that *compound options*[8] are computationally hard to price correctly. These pieces of work study types of contracts that are "complex" even in isolation. In contrast, a single CDS is a very simple contract.[9] Hence, we show in this paper that otherwise simple derivatives, if they occur as part of an otherwise simple network

---

"continues to far exceed the stock of corporate bonds and loans on which most contracts are written." Crotty (2009) quotes Eric Dinallo, then Superintendent of Insurance for New York State, saying that 80 percent of the CDSs outstanding are speculative (i.e., naked). Regulatory changes after 2009, like central clearing and portfolio compression (see Section 7), may have reduced the share of naked CDSs, but they cannot eliminate it below a significant level.

[7]Over-the-counter, i.e., traded directly with other banks rather than through an exchange. In this paper we only consider OTC derivatives.

[8]An option is a derivative that grants the holder the right to buy (*call option*) or sell (*put option*) an asset $A$ at a specified time in the future for a previously agreed-upon price $K$. A compound option is an option where $A$ is itself an option.

[9]Valuation of a CDS is straightforward if distributions of recovery rates are known for the reference entity and counterparty. See Duffie (1999).

structure, create a financial system of high (computational) complexity.

The only other computational complexity result for financial networks we are aware of is by Hemenway and Khanna (2016), who studied the clearing model by Elliott, Golub and Jackson (2014). The authors showed that it is computationally hard to determine the distribution of a given total negative shock to the banks that does the worst damage in terms of value. In contrast, we prove in this work that in financial networks with CDSs, it is already hard to determine the impact of a *known* distribution of shocks to banks. To the best of our knowledge, we are the first to present a computational hardness result for the clearing problem.

### Techniques Used

Since the clearing problem refers to an explicit network, it is natural for us to employ reduction from circuit problems to prove our hardness results.

To prove that deciding existence of a solution is NP-hard, we perform reduction from the CIRCUIT SATISFIABILITY problem. We encode Boolean circuits in a way reminiscent of electrical circuits. Boolean values are represented by recovery rates that are bounded away from $1/2$ by a constant and we define two *financial system gadgets*: one that allows a bank to have either a low or a high recovery rate, for the inputs, and one that implements a NAND operation, for the gates. We prevent accumulation of errors via a special *reset gadget* that maps low values to $0 \pm \varepsilon$ high values to $1 \pm \varepsilon$. Finally, we add a financial sub-network that has no solution *iff* the recovery rate of the "output bank" of the circuit is low.

We frame the decision problem as a *promise problem*: algorithms are only required to show any useful behavior on "clear-cut" instances where either an exact solution exists or not even an $\varepsilon$-approximate solution exists, for some small $\varepsilon$. For intermediate instances (*only* an approximate solution exists), any behavior including non-termination is acceptable. Promise problems are useful for problems where solutions may not be of polynomial length. For example, Schoenebeck and Vadhan (2012) used this approach in the context of certain classes of Nash equilibria. See Goldreich (2005) for a further discussion.

We show PPAD-hardness of the search problem via reduction from generalized circuits. Originally developed for the analysis of the complexity of finding a Nash equilibrium (Daskalakis, Goldberg and Papadimitriou, 2009; Chen, Deng and Teng, 2009; Rubinstein, 2018), generalized circuits have found application in the study of other total search problems. A generalized circuit consists of arithmetic gates and comparison gates,[10] and it can have cycles. The associated search problem asks for a vector of values that is approximately consistent with each gate. Our reduction is straightforward: for each type of gate, we define a gadget that (approximately) performs the respective operation on the recovery rates. PPAD-hardness then follows from hardness of generalized circuits for constant $\varepsilon$ (Rubinstein, 2018). Hardness for *constant $\varepsilon$* is the strongest kind of hardness result one can obtain here and precludes existence of a PTAS unless P=PPAD. If $\varepsilon$ shrinks polynomially as the input grows, this only precludes an FPTAS. If it shrinks exponentially, it only precludes membership in P.

---

[10]Traditionally, generalized circuits have also supported Boolean gates that operate on approximate Boolean values similar to above. Schuldenzucker and Seuken (2019) recently showed that the Boolean gates are in fact redundant and can therefore be omitted.

To show that already finding a set of defaulting banks is PPAD-hard, we define a new discrete variant of the generalized circuit problem, which may be of independent interest. In this problem, we only ask for one of three *states* for each gate: *high*, *medium*, or *low*. These states correspond to "decision" or "truncation points" in the definition of the gates. For example, the addition gate is in a *high* state if and only if its inputs sum to more than one and its output is therefore truncated at one. States also allow for $\varepsilon$ errors. It is PPAD-complete to find a collection of states consistent with some $\varepsilon$-solution of the circuit because with states fixed, the constraints on the gates are linear and one can reconstruct an $\varepsilon$-solution via linear programming. We then show that in our above reduction from financial networks to generalized circuits, the set of defaulting banks already determines the states of the gates. We hope that our technique may be useful to prove hardness of discrete versions of other search problems in the future.

**Basic Notation**

Throughout this paper, we say that $\varepsilon$ is *sufficiently small* (in formulas: $\varepsilon \ll 1$) if it is below a certain positive threshold, where the exact value of the threshold is not relevant in the following. The threshold may depend on parameters that are arbitrary, but fixed, but it will *never* depend on the input to any computational problem and should therefore treated as a constant. We sometimes write $\varepsilon \ll \beta$ to indicate that the threshold is a monotonic (not necessarily linear) function of a term $\beta$. We write $\Theta(\varepsilon)$ for $c\varepsilon$ where $c > 0$ is a certain constant that is not relevant in the following and does not depend on any values or parameters. We define $[x] := \min(1, \max(0, x))$, the *truncation* of $x$ to the interval $[0,1]$. We write $x = y \pm \varepsilon$ for $|x - y| \le \varepsilon$ if $x$ and $y$ are scalars and for $\|x - y\|_\infty \le \varepsilon$ if they are vectors. We also use the notation "$\pm\varepsilon$" in compound expression like $[x \pm \varepsilon]$ to indicate a range of values. This notation formally corresponds to interval arithmetic.

## 2 Financial Networks with Credit Default Swaps[11]

Throughout this paper, we use our clearing model from Schuldenzucker, Seuken and Battiston (2019). We now present the details of this model and provide a high-level discussion of the effects of CDSs. We then present a new relaxation of our solution concept, which is necessary to be able to receive an (approximate) solution of finite, polynomial length.

### 2.1 The Model

**Financial System.** Let $N$ denote a finite set of *banks*. Each bank $i \in N$ holds a certain amount of *external assets*, denoted by $e_i \geq 0$. Between any two banks $i$ and $j$, $|N| + 1$ numbers capture the contracts from the contract *writer* $i$ to the *holder* $j$. Let $c_{i,j}^{\emptyset} \geq 0$ be the total amount of debt that $i$ owes to $j$ and for $k \in N$ let $c_{i,j}^k \geq 0$ be the total amount of CDSs from $i$ to $j$ with *reference entity* $k$. We also call the numbers $c_{i,j}^k$ ($k \in N \cup \{\emptyset\}$) the *notionals* of the respective contracts. If $c_{i,j}^k > 0$ for some $k \in N \cup \{\emptyset\}$, we call $j$ a *creditor* of $i$ and $j$ a *debtor* of $i$.

We make two *sanity assumptions* to rule out pathological cases. First, no bank may enter into a contract with itself or on itself (i.e., $c_{i,i}^{\emptyset} = c_{i,i}^j = c_{i,j}^j = c_{i,j}^i = 0$ for all $i,j \in N$). Second, as CDSs are defined as insurance on debt, we require that any bank that is a reference entity in a CDS must also be writer of some debt contract (i.e., if $\sum_{k,l \in N} c_{k,l}^i > 0$, then $\sum_{j \in N} c_{i,j}^{\emptyset} > 0$, for all $i \in N$).[12]

We model default costs following Rogers and Veraart (2013): there are two *default cost parameters* $\alpha, \beta \in [0,1]$. Defaulting banks are only able to pay to their creditors a share of $\alpha$ of their external assets and a share of $\beta$ of their incoming payments. Thus, $\alpha = \beta = 1$ means that there are no default costs and $\alpha = \beta = 0$ means that assets held by defaulting banks are worthless. The values $1 - \alpha$ and $1 - \beta$ are the *default costs*.

A *financial system* is a tuple $(N, e, c, \alpha, \beta)$ where $N$ is a set of banks, $e$ is a vector of external assets, $c$ is a 3-dimensional matrix of contracts, and $\alpha$ and $\beta$ are default cost parameters. Note that, even though $\alpha$ and $\beta$ are part of the definition of a financial system, our results in this paper will be for restrictions of the respective problems to arbitrary but fixed values of $\alpha$ and $\beta$. The parameters can therefore be considered constant.

Note that we neither specify a distribution of shocks nor an initial payment when a contract is made. We assume that these values are implicitly reflected in the external

---

[11]An extended version of subsections 2.1, 2.2, and parts of Effects of Allowing CDSs of this section has previously appeared in our prior work (Schuldenzucker, Seuken and Battiston, 2019, Section 2). We repeat a short version here for convenience. See the aforementioned paper for a discussion of alternative models for financial networks with CDSs.

[12]For technical reasons, we allow our *financial system gadgets* in section 3 and 4 to violate the second assumption. In this case, the violating banks will be "dummy banks" that hold and write no contracts and are ignored when considering solutions.

assets.

**Assets and Liabilities.** We are ultimately looking for a vector of *recovery rates* $r_i \in [0, 1]$. For any two banks banks $i$ and $j$, the contracts from $i$ to $j$ give rise to a *liability* of $i$ to $j$. This is the amount of money that $i$ has to pay to $j$. A debt contract gives rise to an unconditional liability equal to its notional, but the liability in a CDS on some bank $k$ depends on the recovery rate of $k$ and is proportional to $1 - r_k$. The total liability from $i$ to $j$ at $r$ is therefore:

$$l_{i,j}(r) := c_{i,j}^{\emptyset} + \sum_{k \in N} (1 - r_k) \cdot c_{i,j}^k$$

The *total liabilities* of $i$ at $r$ are the aggregate liabilities that $i$ has toward all other banks, denoted by

$$l_i(r) := \sum_{j \in N} l_{i,j}(r).$$

The actual *payment* $p_{i,j}(r)$ from $i$ to $j$ at $r$ can be lower than $l_{i,j}(r)$ if $i$ is in default. A bank that is in default makes payments for its contracts in proportion to the respective liability;

$$p_{i,j}(r) := r_i \cdot l_{i,j}(r).$$

The *total assets* $a_i(r)$ of a bank $i$ at $r$ consist of its external assets $e_i$ and the incoming payments;

$$a_i(r) := e_i + \sum_{j \in N} p_{j,i}(r).$$

In case bank $i$ is in default, its *assets after default costs* $a_i'(r)$ are the assets reduced according to the factors $\alpha$ and $\beta$. This is the amount that will be paid out to creditors;

$$a_i'(r) := \alpha e_i + \beta \sum_{j \in N} p_{j,i}(r).$$

*Remark* 1. To see that CDSs indeed act as insurance on default, let $i, j, k \in N$ and assume that bank $j$ holds both debt from $k$ and a CDS on $k$, both with the same notional: $c_{i,j}^k = c_{k,j}^{\emptyset} =: \delta > 0$. Then the assets of $j$ contain the term $r_k \cdot \delta + r_i \cdot (1 - r_k) \cdot \delta$. As long as $r_i = 1$, this term is equal to $\delta$ independently of $r_k$.

**Clearing Recovery Rate Vector.** Following Eisenberg and Noe (2001), we call a recovery rate vector $r \in [0, 1]^N$ *clearing* if it satisfies the essential principles of bankruptcy law:

1. Banks with sufficient assets to pay their liabilities in full must do so.

2. Banks with insufficient assets to pay their liabilities in full are in default and must pay out all their assets to creditors after default costs have been subtracted.

This leads to the following formal definition:

**Figure 1** Example financial system. Let $\alpha = \beta = 0.5$

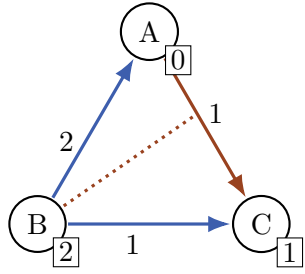

**Definition 1** (Clearing Recovery Rate Vector)**.** Let $X = (N, e, c, \alpha, \beta)$ be a financial system. Define the *update function*

$$F : [0,1]^N \to [0,1]^N$$

$$F_i(r) := \begin{cases} 1 & \text{if } a_i(r) \geq l_i(r) \\ \frac{a'_i(r)}{l_i(r)} & \text{if } a_i(r) < l_i(r). \end{cases}$$

A recovery rate vector $r \in [0,1]^N$ is called *clearing* for $X$ if it is a fixed point of the update function, i.e., if $F_i(r) = r_i$ for all $i$. We also call a clearing recovery rate vector a *solution* to the clearing problem.

## 2.2 Example and Visual Representation

Figure 1 shows a visual representation of an example financial system. There are three banks $N = \{A, B, C\}$, drawn as circles, with external assets of $e_A = 0$, $e_B = 2$, and $e_C = 1$, drawn as rectangles on top of the banks. Debt contracts are drawn as blue arrows from the writer to the holder and they are annotated with the notionals $c_{B,A}^\emptyset = 2$ and $c_{B,C}^\emptyset = 1$. CDSs are drawn as orange arrows, where a dashed line connects to the reference entity, and are also annotated with the notionals: $c_{A,C}^B = 1$. Default cost parameters $\alpha = \beta = 0.5$ are given in addition to the picture.

A clearing recovery rate vector for this example is given by $r_A = 1$, $r_B = \frac{1}{3}$, and $r_C = 1$. The liabilities arising from this recovery rate vector are $l_{B,A}(r) = 2$, $l_{B,C}(r) = 1$, and $l_{A,C}(r) = \frac{2}{3}$. Payments are $p_{B,A}(r) = \frac{2}{3}$, $p_{B,C} = \frac{1}{3}$, and $p_{A,C}(r) = \frac{2}{3}$. This is the only solution for this system.

## 2.3 Effects of Allowing CDSs

The addition of CDSs significantly changes the mathematical properties of the model. Most importantly, the assets of a bank $i$ can now contain terms of form $c_{j,i}^k \cdot r_j \cdot (1 - r_k)$ and are thus *non-linear* and *non-monotonic* in the recovery rates of the other banks, and so is the update function $F$. In contrast, if only debt contracts are allowed, the assets are always linear and monotonic and $F$ is piecewise linear and monotonic, where the linear segments are given by regions where the set of defaulting banks does not change. Eisenberg and Noe (2001) exploited this for their *fictitious default algorithm*: we keep track of a candidate set of defaulting banks, beginning with the empty set. In each step, we solve a linear equation system to compute clearing recovery rates assuming this set of banks defaults. We then update the set of defaulting banks. By monotonicity, this set can only grow over time and we terminate at a solution after at most $|N|$ steps. In particular, a rational solution of polynomial length always exists. If there are default costs, a discontinuity appears at the boundary of the default regions, but, as Rogers and Veraart (2013) have shown, the algorithm still works.

The fictitious default algorithm clearly does not extend to CDSs. First, since the update function is now not necessarily linear (and also not convex or concave), the individual steps cannot easily be performed in polynomial time. Further, by non-monotonicity, the set of defaulting banks need not grow monotonically over time and thus the algorithm would not necessarily terminate. Indeed, we show in Section 5 that already finding the correct set of defaulting banks is hard.

In Schuldenzucker, Seuken and Battiston (2019) we showed that the combination of non-monotonicity and the discontinuity introduced by default costs can create a situation where no solution to the clearing problem exists. The idea of our counterexample was that if a bank holds a CDS on itself, then this can lead to a situation where default of this bank implies non-default and vice versa. In a network, an equivalent situation can arise indirectly. The discontinuity ensures that a "middle ground" is not attainable. We present a generalized construction for this kind of system in Section 3 below.

We have considered two important special cases in our above prior work. First, if there are no default costs ($\alpha = \beta = 1$), continuity is restored[13] and one can show using a fixed-point theorem that a solution always exists. In Section 4 of this paper we show, however, that *finding* (an approximation of) this solution is hard. The second special case is when *naked CDSs* are not allowed and where *monotonicity* is restored. In this case, we have received a *constructive* proof of existence and we will show in Section 6.2 that it in fact constitutes an FPTAS.

### 2.4 Approximate Solutions

Due to the non-linearity discussed above, there are financial systems where all solutions contain irrational numbers (see Appendix A for an example). To receive finite, polynomial-length objects, we relax Definition 1 to an approximate solution as follows.

**Definition 2** (Approximately Clearing Recovery Rate Vector)**.** Let $X = (N, e, c, \alpha, \beta)$ be a financial system and let $\varepsilon \geq 0$. A recovery rate vector $r$ is called $\varepsilon$-*approximately*

---

[13]The function $F$ can still contain a discontinuity at the boundaries of the sets $\{r \mid l_i(r) = 0\}$. This is easy to circumvent, though.

*clearing* (or an *ε-solution*) for $X$ if for each $i \in N$ at least one of the following two conditions is satisfied:

$$r_i = 1 \pm \varepsilon \text{ and } a_i(r) \geq (1 - \varepsilon)l_i(r)$$

$$r_i = \frac{a'_i(r)}{l_i(r)} \pm \varepsilon \text{ and } a_i(r) < (1 + \varepsilon)l_i(r)$$

Note how both the "case selection part" and the "output part" of the definition of the function $F$ are relaxed. Banks with much higher assets than liabilities are unambiguously not in default and those with much lower assets than liabilities are in default. But when assets approximately equal liabilities, a bank can be considered either of the two. Default in the real world is not a knife-edge decision, but has some tolerance. Therefore, the above definition likely reflects what (say) a regulator running a stress test will be interested in. Recall that the precision $\varepsilon$ is defined in the space of *recovery rates*. That is why $\varepsilon$ is used as an additive error in the output recovery rate, but a multiplicative one when comparing assets and liabilities.

We take the following elementary properties as evidence that our approximate solution concept is natural. The proof is straightforward and hence omitted.

**Proposition 1.** *Let $X = (N, e, c, \alpha, \beta)$ be a financial system.*

1. *An exact solution is the same as a 0-solution. If $\varepsilon' > \varepsilon \geq 0$, then any $\varepsilon$-solution is also an $\varepsilon'$-solution.*

2. *If $F(r) = r \pm \varepsilon$, then $r$ is an $\varepsilon$-solution.*

3. *If $\alpha = \beta = 1$, then $F(r) = r \pm \varepsilon$ if and only if $r$ is an $\varepsilon$-solution.*

4. *If $r$ is an $\varepsilon$-solution, then we have:*

$$a_i(r) \geq (1 + \varepsilon)l_i(r) \Rightarrow r_i = 1 \pm \varepsilon$$

$$a_i(r) < (1 - \varepsilon)l_i(r) \Rightarrow r_i = \frac{a'_i(r)}{l_i(r)} \pm \varepsilon$$

5. *If $r$ is an $\varepsilon$-solution and $l_i(r) > 0$, then $r_i \leq \frac{a_i(r)}{l_i(r)} + \varepsilon$.*

6. *If $r$ is an $\varepsilon$-solution and $r_i < 1 - \varepsilon$, then $r_i \leq \max(\alpha, \beta) + \varepsilon$. If in addition $e_i = 0$, then $r_i \leq \beta + \varepsilon$.*

Throughout this paper, we will make the following additional technical assumption.

**Definition 3** (Non-degenerate Financial System)**.** A financial system $X = (N, e, c, \alpha, \beta)$ is called *non-degenerate* if every bank that writes any contracts also writes a debt contract. That is, for all $i \in N$, if $\sum_{j,k \in N} c_{i,j}^k > 0$, then $\sum_{j \in N} c_{i,j}^{\emptyset} > 0$.

Note that non-degeneracy is a very weak requirement in the real world. All we demand is that every bank has *some* constant liabilities, for example to its customers.

In non-degenerate financial systems, we can round any exact solution to receive an $\varepsilon$-solution of polynomial length.

**Lemma 1.** *Let $\varepsilon > 0$ and let $X$ be a non-degenerate financial system. If $X$ has an exact solution, then $X$ has an $\varepsilon$-solution of size polynomial in the sizes of $\varepsilon$ and $X$.*

*Proof.* Assume WLOG that every bank writes a debt contract. If this is not true for some bank, no other bank depends on its recovery rate by non-degeneracy and our sanity assumptions. We can thus simply set its recovery rate to 1.

Note that the functions $\frac{a_i}{l_i}$ and $\frac{a'_i}{l_i}$ are *polynomially continuous* in $X$, i.e., continuous with a Lipschitz constant that is $O\left(2^{\text{poly}(\text{size}(X))}\right)$. This is because $a_i$, $a'_i$, and $l_i$ are polynomially continuous and $l_i$ is bounded above $\sum_j c^{\emptyset}_{i,j} > 0$ because every bank writes a debt contract. In particular, $\frac{a_i(r)}{l_i(r)}$ and $\frac{a'_i(r)}{l_i(r)}$ is well-defined for all $r$. Let $M$ be the maximum of the Lipschitz constants of these functions and 1.

Let $r$ be an exact solution and let $r'$ be $r$ rounded to a multiple of $\delta := \varepsilon/(M+1)$, so that $r' = r \pm \delta$. By polynomial continuity, $r'$ has a size as required. To see that $r'$ is an $\varepsilon$-solution, we perform a case distinction for each $i$:

- If $r_i = 1$, then $r'$ satisfies the first case in Definition 2. We have $r'_i = r_i \pm \delta = 1 \pm \varepsilon$. Further, since $r_i = 1$ we have $\frac{a_i(r)}{l_i(r)} \geq 1$, by choice of $M$ and $r'$, $\frac{a_i(r')}{l_i(r')} = \frac{a_i(r)}{l_i(r)} \pm M\delta \geq 1 - \varepsilon$, and thus $a_i(r') \geq (1-\varepsilon)l_i(r')$.

- If $r_i < 1$, then $r'$ satisfies the second case. We have $r'_i = r_i \pm \delta = \frac{a'_i(r)}{l_i(r')} \pm \delta = \frac{a'_i(r')}{l_i(r')} \pm (M+1)\delta = \frac{a'_i(r')}{l_i(r')} \pm \varepsilon$. Since $r_i < 1$, we have $\frac{a_i(r)}{l_i(r)} < 1$ and thus $\frac{a_i(r')}{l_i(r')} = \frac{a_i(r)}{l_i(r)} \pm \varepsilon < 1 + \varepsilon$, that is, $a_i(r') < (1+\varepsilon)l_i(r')$. □

The proof of the previous lemma made use of the fact that we relaxed both sides of the definition of $F$ in our definition of an $\varepsilon$-solution. Note that, since $F$ is not usually continuous, we do not necessarily receive an approximate fixed point of $F$ via rounding. Note further that the lemma does not imply that *all* $\varepsilon$-solutions are close to an exact solution (if one exists) or have polynomial length. This is common for approximate solution concepts.[14]

---

[14]Using a theorem by Anderson (1986), it follows from the syntactic structure of the definition that for any $X$ and $\delta$ there is an $\varepsilon$ such that any $\varepsilon$-solution for $X$ is $\delta$-close to an $r$ that is *almost* an exact solution: $r_i = F_i(r)$ *unless* $a_i(r) = l_i(r)$, in which case we can have $r_i = 1$ or $r_i = a'_i(r)/l_i(r) = a'_i(r)/a_i(r)$. However, $\varepsilon$ depends on $X$ in this case, while we consider a constant $\varepsilon$ in this paper. It is easy to construct examples where for constant $\varepsilon$, exact and approximate solutions are far apart. Etessami and Yannakakis (2010) studied search problems for *strong* approximate fixed points, which need to be close to an exact fixed point. Computational complexity is markedly higher than for the regular approximate variants. For example, it is an open question if finding a strong approximate Nash equilibrium is even in NP.

# 3  The Complexity of Deciding Existence of a Solution

The goal of this section is to prove our first main result: It is hard to distinguish between financial systems that have an exact solution and those that have no $\varepsilon$-solution.

**Theorem 1.** *For any fixed $\alpha$ and $\beta$ such that $\alpha < 1$ or $\beta < 1$ and for $\varepsilon$ sufficiently small depending on $\alpha$ and $\beta$, the promise problem $\varepsilon$-HasClearing$^{\alpha,\beta}$, defined as follows, is NP-complete: Given a non-degenerate financial system $X = (N, e, c, \alpha, \beta)$,* ...

- *if $X$ has an exact solution, return* Yes.

- *if $X$ has no $\varepsilon$-solution, return* No.

If none of the two conditions is satisfied, any behavior including non-termination is allowed. Membership in NP follows from Lemma 1 because it is enough to decide if an $\varepsilon$-solution of a certain polynomial maximum length exists. One might argue that it would be more natural to instead consider the problem "Given $X$, return Yes if an $\varepsilon$-solution exists and No otherwise." Theorem 1 implies that this problem is NP-hard for $0 \leq \varepsilon \ll 1$, but it does not imply that it is a member of NP. This is because Lemma 1 does *not* imply that there is *always* an $\varepsilon$-solution of polynomial length if an $\varepsilon$-solution exists, but only when an exact solution exists.[15] Note further that $\varepsilon$-HasClearing$^{\alpha,\beta}$ becomes (weakly) easier as we increase $\varepsilon$, while this is not clear for the above non-promise variant. Framing the problem as a promise problem avoids these issues.

The remainder of this section is dedicated to showing hardness of the $\varepsilon$-HasClearing$^{\alpha,\beta}$ problem. Our reduction is from Circuit Satisfiability. We will represent Boolean values by recovery rates that are contain in the set $[0, 1/4] \cup [3/4, 1]$, with the low part of this set representing False and the high part representing True. We then encode the inputs and the gates using *financial system gadgets* and we force the output to True by adding another special financial sub-system. We prevent error accumulation using a special *reset gadget* not unlike the *brittle comparison* gadget in Daskalakis, Goldberg and Papadimitriou (2009).

## 3.1  Financial System Gadgets

We first introduce some technical machinery for the proofs in this section and Section 4. We define a *financial system gadget* as a small financial system where the recovery rate of an *output bank* depends on a collection of *input banks* in a certain way. We can "plug" a gadget into another financial system by simply identifying its input banks with some other banks. We will use this to successively build up financial systems.

---

[15]The proof of the lemma implies that an $\varepsilon$-solution of length polynomial in the length of $\varepsilon - \varepsilon_0$ exists if an $\varepsilon_0$-solution exists, for some $0 \leq \varepsilon_0 < \varepsilon$. To derive membership in NP for the non-promise problem, we would need to bound $\inf \{\varepsilon_0 \mid \text{an } \varepsilon_0\text{-solution exists}\}$ from above relative to the length of $X$. We leave it to future work to explore to which extent this might be possible.

**Definition 4** (Financial System Extension and Gadget). An *extension* of a financial system $X = (N, e, c, \alpha, \beta)$ is a financial system $X' = (N', e', c', \alpha, \beta)$ such that $N \subseteq N'$ and the assets and liabilities of each bank $i \in N$ are the same in $X$ and $X'$. That is, we have $e'_i = e_i \ \forall i \in N$ and if $c'^k_{i,j} > 0$, then either i) $i, j \notin N$ or ii) $i, j \in N$, $k \in N \cup \{\emptyset\}$, and $c'^k_{i,j} = c^k_{i,j}$. We call $X'$ an *extension* of $X$ *on* $N_0 \subseteq N$ if this holds for all $i, j \in N_0$. We call $r \in [0, 1]^N$ an $\varepsilon$-solution *on* $N_0$ if the condition from Definition 2 holds for all $i \in N_0$.

A *financial system gadget* is a financial system $G = (N, e, c, \alpha, \beta)$ with a set of distinguished *input banks* $A := \{a_1, \ldots, a_m\} \subseteq N$ that have no assets or liabilities (i.e., $e_{a_i} = c^k_{a_i, j} = c^k_{j, a_i} = 0$ for $i = 1, \ldots, m$, $j \in N$, and $k \in N \cup \{\emptyset\}$) such that the following property holds: for any $r_A \in [0, 1]^A$ there exists an $r_{N \setminus A} \in [0, 1]^{N \setminus A}$ such that $r_A \cup r_{N \setminus A}$ is an exact solution on $N \setminus A$. We say that $G$ *implements* a property $P : [0, 1]^N \to \{\text{TRUE}, \text{FALSE}\}$ if for any sufficiently small $\varepsilon$ and any $\varepsilon$-solution $r$ of $G$ on $N \setminus A$, $P(r)$ holds. If $X$ is a financial system and $a'_1, \ldots, a'_m$ are banks in $X$ each of which writes a debt contract, then the *application* of $G$ to $X$ and $a'_1, \ldots, a'_m$ is a new financial system $X'$ obtained as the union of $X$ and $G$ where we identify $a_i$ and $a'_i$ for $i = 1, \ldots, m$. Note that $X'$ is an extension of $X$ and an extension of $G$ on $N \setminus A$.

*Remark* 2 (Applying Several Gadgets). Gadgets are compatible with modifications of the financial system and in particular with each other. To see this, first note that if $X'$ is an extension of $X$ on $N_0$ and $r'$ is an $\varepsilon$-solution for $X'$, then the restriction $r'|_N$ is an $\varepsilon$-solution for $X$ on $N_0$. Assume now that a gadget $G = (N, e, c, \alpha, \beta)$ with inputs $A$ implements a property $P$ and $X''$ is an extension of $G$ on $N \setminus A$. $X''$ could result, for example, from applying $G$ to some financial system and then applying arbitrary other gadgets on top. Let $\varepsilon \ll 1$ and let $r$ be an $\varepsilon$-solution for $X''$. Then, by the extension, $r|_N$ is an $\varepsilon$-solution for $G$ on $N \setminus A$ and thus, $P(r|_N)$ holds.

Our financial system gadgets will have between zero and two input banks, which we will call $a$ and $b$ for convenience. They will also have an *output bank*, called $v$, and they will implement properties that make the recovery rate of the output bank equal to a certain function of the recovery rates of the input banks, up to errors. We will assume that each gadget contains a *source bank* $s$ that holds no contracts and a *sink bank* $t$ that writes no contracts. The other banks hold CDSs from $s$ and write a debt contract of notional 1 to $t$, so that $l_i(r) = 1$. The connection to the inputs is established via CDS references. We set $c^\emptyset_{s,t} = 1$ to ensure non-degeneracy and we further always set $e_s \geq 2 \sum_{i \in N, k \in N \cup \{\emptyset\}} c^k_{s,i}$. This implies that $s$ cannot default and thus $r_s, r_t = 1 \pm \varepsilon$ in any $\varepsilon$-solution. For the sake of conciseness, we will leave out these contracts and external assets in our descriptions of the gadgets. We mark the input banks by dashed outlines in our figures.

**Figure 2** Zero-One Gadget. Variant of Schuldenzucker, Seuken and Battiston (2019, Figure 3).

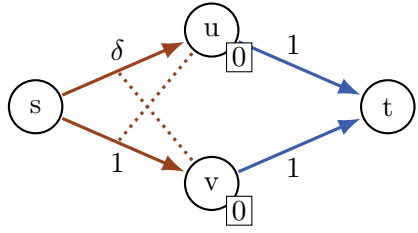

## 3.2 Reducing Boolean Circuits to Finacial Systems

We begin with a financial system gadget where the output bank can have recovery rates approximately 0 or 1. We will use this to encode variables.

**Lemma 2** (Zero-One Gadget). *For all $\alpha, \beta \in [0,1]$ there is a financial system gadget with no input banks such that the following hold:*

- *There exists an exact solution where $r_v = 0$ and one where $r_v = 1$.*

- *The gadget implements the following property: $r_v = 0 \pm \Theta(\varepsilon)$ or $r_v = 1 \pm \Theta(\varepsilon)$.*

*Proof.* Consider the financial system in Figure 2. We distinguish the cases $\beta < 1$ and $\beta = 1$.

If $\beta < 1$, let $\delta = 2\frac{1}{1-\beta}$. It is easy to see that $(r_u, r_v) \in \{(0,1),(1,0)\}$ are exact solutions. We show that in any $\varepsilon$-solution $r$ we have $r_v = 1 \pm \varepsilon$ or $r_v = 0 \pm 2\varepsilon$. To see this, let $r_v < 1 - \varepsilon$. Then $r_v \leq \beta + \varepsilon$ by Proposition 1 and thus $a_u(r) = \delta r_s(1 - r_v) \geq \delta(1-\varepsilon)(1-\beta-\varepsilon) = 2(1-\varepsilon)(1-\frac{\varepsilon}{1-\beta}) \geq 1+\varepsilon = (1+\varepsilon)l_u(r)$ where the last inequality holds for $\varepsilon \ll 1 - \beta$. Thus, $r_u = 1 \pm \varepsilon$ and thus $a_v(r) \leq \varepsilon$, so $r_v \leq \frac{a_v(r)}{l_v(r)} + \varepsilon = 2\varepsilon$.

If $\beta = 1$, let $\delta = 2$. It is again easy to see that $(r_u, r_v) \in \{(0,1),(1,0)\}$ are exact solutions. Let $r$ be an $\varepsilon$-solution. For $i = u, v$ we have $a_i(r) = a_i'(r) \ \forall r$. This follows from the definition of $a_i'$ because $\beta = 1$ and $e_i = 0$. Like in Proposition 1 part 3, this implies $r_i = F_i(r) \pm \varepsilon = [a_i(r)] \pm \varepsilon$. That is:

$$r_v = [(1 \pm \varepsilon)(1 - r_u)] \pm \varepsilon = 1 - r_u \pm 2\varepsilon$$

$$r_u = [2(1 \pm \varepsilon)(1 - r_v)] \pm \varepsilon = [2(1 - r_v)] \pm 3\varepsilon$$

Taken together, these imply:

$$r_v = 1 - [2(1 - r_v)] \pm 5\varepsilon = [1 - 2(1 - r_v)] \pm 5\varepsilon = [2r_v - 1] \pm 5\varepsilon \qquad (*)$$

We now perform a case distinction on $r_v$.

- If $r_v \geq 1/2$, then $[2r_v - 1] = 2r_v - 1$ and thus by $(*)$, $r_v = 1 \pm 5\varepsilon$.

- If $r_v < 1/2$, then $[2r_v - 1] = 0$ and thus by $(*)$, $r_v = 0 \pm 5\varepsilon$. $\qquad\square$

**Figure 3** Cutoff Gadget

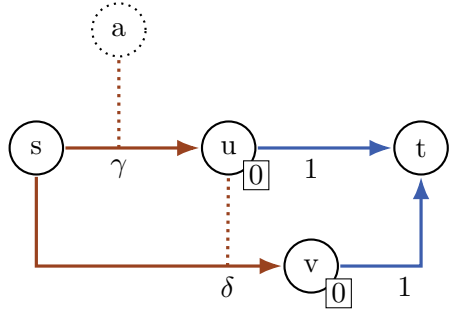

We next work towards a NAND gadget to encode Boolean gates. We begin by introducing a versatile tool that can be used to map values significantly above or below certain thresholds to approximately 0 or 1. We will use this *cutoff gadget* in this section and Section 4.

**Lemma 3** (Cutoff Gadget). *Let $0 < K < L < 1$. There exists a financial system gadget with one input bank that implements the following property for all $\alpha, \beta \in [0, 1]$:*

$$r_a \leq K - \Theta\left(\varepsilon\right) \Rightarrow r_v = 0 \pm \Theta\left(\frac{\varepsilon}{L-K}\right)$$

$$r_a \geq L + \Theta\left(\varepsilon\right) \Rightarrow r_v = 1 \pm \Theta\left(\frac{\varepsilon}{L-K}\right).$$

*Proof.* Consider the gadget in Figure 3[16] where we set:

$$\gamma := \frac{1}{1-K}$$

$$\delta := \frac{1-K}{L-K}$$

Consider an $\varepsilon$-solution. Let first $r_a \leq K - 3\varepsilon$. Then $a_u(r) = \gamma r_s(1 - r_a) \geq \gamma(1-\varepsilon)(1 - K + 3\varepsilon) = (1-\varepsilon)(1 + \frac{3\varepsilon}{1-K}) \geq (1-\varepsilon)(1 + 3\varepsilon) = 1 + 2\varepsilon - 3\varepsilon^2 \geq 1 + \varepsilon$ if $\varepsilon \ll 1$. Thus, $r_u = 1 \pm \varepsilon$. Now $a_v(r) = \delta r_s(1 - r_u) \leq \delta\varepsilon \leq \frac{2\varepsilon}{L-K}$. And $r_v \leq \frac{a_v(r)}{l_v(r)} + \varepsilon = a_v(r) + \varepsilon \leq a_v(r) + \frac{3\varepsilon}{L-K}$.

Let now $r_a \geq L + 4\varepsilon$. Then $a_u(r) \leq \gamma(1 - L - 4\varepsilon) = \frac{1-L}{1-K} - \frac{4\varepsilon}{1-K}$ and thus $r_u \leq \frac{1-L}{1-K} - \frac{4\varepsilon}{1-K} + \varepsilon$. Now $a_v(r) \geq \delta(1-\varepsilon)\left(1 - \frac{1-L}{1-K} + \frac{4\varepsilon}{1-K} - \varepsilon\right) = (1-\varepsilon)\left(1 + \frac{(3+K)\varepsilon}{L-K}\right) \geq (1-\varepsilon)(1 + 3\varepsilon) = 1 + 2\varepsilon - 3\varepsilon^2 \geq 1 + \varepsilon$ for $\varepsilon \ll 1$. Thus, $r_v = 1 \pm \varepsilon$. □

The cutoff gadget has two kinds of errors. First, there is always a region $r_a \in (K - \Theta(\varepsilon), L + \Theta(\varepsilon)) \supset (K, L)$ where the output value is unspecified and depends on error terms more than on the input. This is reminiscent of the *brittle comparison* gate in generalized circuits (see Section 4). Second, there is an error in the output that depends on how much "brittleness" we are willing to tolerate. We will show in Section 4 how this trade-off can be circumvented. In this section, we use a large brittleness to receive a gadget prevents error accumulation in the Boolean gadgets below.

---

[16]Note that the gadget violates our sanity assumptions because the input banks is a CDS reference entity, but not writers of any debt contracts. In this case, this does not cause any problems. Note in particular that, as soon as the gadget is applied to some other system, the sanity assumption will hold.

**Figure 4** NAND Gadget

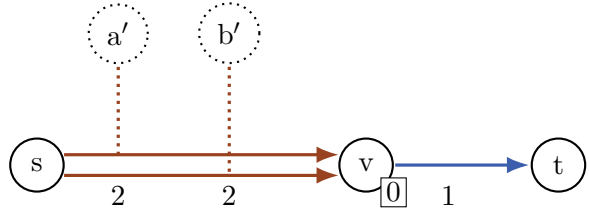

**Corollary 1** (Reset Gadget)**.** *There exists a financial system gadget with one input bank that implements the following property for all $\alpha, \beta \in [0, 1]$:*

$$r_a \leq 1/4 \Rightarrow r_v = 0 \pm \Theta(\varepsilon)$$
$$r_a \geq 3/4 \Rightarrow r_v = 1 \pm \Theta(\varepsilon).$$

*Proof.* Use a cutoff gadget with $K = 2/5$ and $L = 3/5$. For $\varepsilon \ll 1$ we have $1/4 \leq K - \Theta(\varepsilon)$ and $3/4 \geq L + \Theta(\varepsilon)$ and $L - K = 1/5$ is a constant. □

To represent Boolean gates, we introduce a gadget that mirrors the Boolean operation $x$ NAND $y = \neg(x \land y)$.

**Lemma 4** (NAND Gadget)**.** *There exists a financial system gadget with two input banks that implements the following property for all $\alpha, \beta \in [0, 1]$:*

$$r_a \leq 1/4 \ or \ \ r_b \leq 1/4 \Rightarrow r_v = 1 \pm \Theta(\varepsilon)$$
$$r_a \geq 3/4 \ and \ r_b \geq 3/4 \Rightarrow r_v = 0 \pm \Theta(\varepsilon)$$

*Proof.* Apply the reset gadget each to $a$ and $b$ and call the output banks $a'$ and $b'$, respectively. Then apply the gadget in Figure 4.

Assume first that $r_a \leq 1/4$ or $r_b \leq 1/4$. Then $r_{a'}, r_{b'} = 0 \pm \Theta(\varepsilon)$ and thus $a_v(r) \geq 2(1-\varepsilon)(1-\Theta(\varepsilon)) \geq 1+\varepsilon$ if $\varepsilon \ll 1$. Thus, $r_v = 1\pm\varepsilon$. Assume next that $r_a \geq 3/4$ and $r_b \geq 3/4$. Then $r_{a'}, r_{b'} = 1\pm\Theta(\varepsilon)$ and thus $r_v \leq a_v(r) + \varepsilon \leq 4\Theta(\varepsilon)+\varepsilon = \Theta(\varepsilon)$. □

Using the previous lemma, we easily construct gadgets for all Boolean functions. Note in particular that we can chain NAND gadgets without having to worry about error accumulation because $r_v = 1 \pm \Theta(\varepsilon) \Rightarrow r_v \geq 3/4$ if $\varepsilon \ll 1$, and likewise for 0. A chain of NAND gadgets will thus produce the appropriate output if the inputs to the chain are in $[0, 1/4] \cup [3/4, 1]$. This is why we use a reset gadget. We can combine a collection of zero-one and NAND gadgets to represent a Boolean circuit.

**Corollary 2** (Financial Boolean Circuit)**.** *Let $C$ be a Boolean circuit with $m$ inputs. For $\chi \in \{0, 1\}^m$ write $C(\chi) \in \{0, 1\}$ for the value of the output of $C$ given values $\chi$ at the inputs. For any $\alpha, \beta \in [0, 1]$ and $\varepsilon$ sufficiently small there exists a financial system $X = (N, e, c, \alpha, \beta)$ with $m + 1$ distinguished banks $V := \{a_1, \dots, a_m, v\}$ such that the following hold:*

1. *For any assignment $\chi \in \{0, 1\}^m$ there exists an exact solution $r$ such that $r_{a_i} = \chi_i$ for $i = 1, \dots, m$.*

2. *If $r$ is an $\varepsilon$-solution, then $r_i = 0 \pm \Theta(\varepsilon)$ or $r_i = 1 \pm \Theta(\varepsilon)$ for all $i \in V$.*

3. *If $r$ is an $\varepsilon$-solution and $i \in V$, let $\chi_i = 0$ if $r_i \leq 1/4$ and $\chi_i = 1$ if $r_i \geq 3/4$. Then $\chi_v = C(\chi_{a_1}, \dots, \chi_{a_m})$.*

*Proof.* Assume WLOG that $C$ consists only of NAND gates. We will identify the nodes of $C$ with certain banks in the to-be-constructed financial system. We begin our construction with an empty financial system and build it up iteratively. First apply the zero-one gadget (Lemma 2) $m$ times and identify the output banks of these gadgets with the input nodes of $C$. Now iterate over the gates of $C$ in topological order. For each NAND gate connecting two inputs to an output, by the topological order, the inputs are already nodes in the financial system and the output is not. Apply the NAND gadget (Lemma 4) to the inputs and identify the output bank with the output node of the gate.

Property 1 is satisfied by the zero-one gadget, as is Property 2 for $i = a_1, \dots, a_m$. Property 2 for $i = v$ and property 3 follow by induction on the number of gates by the NAND gadgets. $\qquad\square$

### 3.3 Reducing Satisfiability to Existence of a Solution

The final step in our construction is a way to "destroy" certain unwanted $\varepsilon$-solutions. We will use this to remove exactly those solutions that correspond to falsifying assignments of the Boolean circuit in our previous construction, leaving only those corresponding to satisfying assignments, if any. Note that this cannot be done using a financial system gadget because it does not preserve all existing solutions. The approach is otherwise exactly the same, though.

**Lemma 5.** *Let $\alpha, \beta$ be such that $\alpha < 1$ or $\beta < 1$. There exists a financial system $G = (N, e, c, \alpha, \beta)$ with a distinguished input bank $a \in N$ with no assets or liabilities such that the following hold:*

1. *For any $r_v \geq 3/4$, there exists an $r_{N \setminus \{a\}} \in [0, 1]^{N \setminus \{a\}}$ such that $r_a \cup r_{N \setminus \{a\}}$ is an exact solution on $N \setminus \{a\}$.*

**Figure 5** High-level structure of the financial system in Lemma 5, case $\beta < 1$. Gray boxes indicate gadgets with their output banks. A dashed line with a hollow arrow tip connects a bank to a gadget of which it is an input bank. The parameters of the cutoff gadget are chosen such that $\beta < K < \frac{\beta+1}{2} < L < 1$ evenly spaced.

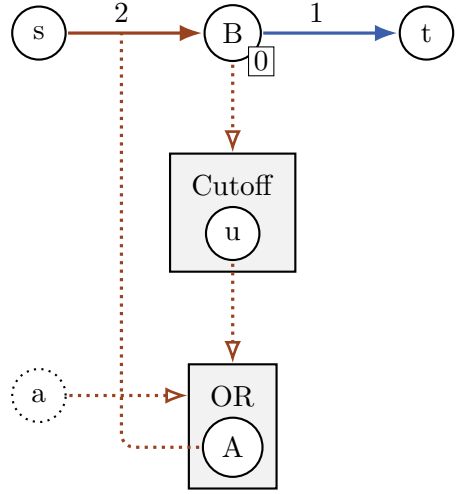

2. *For $\varepsilon \ll 1$, there is no $\varepsilon$-solution $r$ on $N \setminus \{a\}$ where $r_a \leq 1/4$.*

*Proof.* We distinguish the cases $\beta < 1$ and $\alpha < \beta = 1$.

If $\beta < 1$, perform the construction outlined in Figure 5: assume we have a source and a sink bank as usual. Add a new bank $B$ and let $e_B = 0$ and $c_{B,t}^{\emptyset} = 1$. Apply a cutoff gadget to $B$ with $K = \frac{3\beta+1}{4}$ and $L = \frac{\beta+3}{4}$. Note that $\beta < K < \frac{\beta+1}{2} < L < 1$ evenly spaced and the output error of the cutoff gadget is $\Theta(\frac{\varepsilon}{L-K}) = \Theta(\frac{\varepsilon}{1-\beta})$. Call the output bank of the cutoff gadget $u$, apply an OR gadget to $a$ and $u$ and call the output $A$. Finally, add CDS $c_{s,B}^A = 2$.

Towards property 1, if $r_a \geq 3/4$, then by the OR gadget (and this by the NAND gadget), we can set $r_A = 1$ independently of $r_u$. We can then extend $r$ to the other banks via $r_B = 0$ and by setting the recovery rates for the intermediate nodes of the gadgets accordingly.

Towards property 2, if $r_a \leq 1/4$, assume towards a contradiction that $r$ is an $\varepsilon$-solution on $N \setminus \{a\}$. We perform case distinction on $r_B$.

- If $r_B \geq 1 - \varepsilon$, then in particular $r_B \geq L + \Theta(\frac{\varepsilon}{1-\beta})$ if $\varepsilon \ll 1 - \beta$.[17] Thus, by the cutoff gadget and the OR gadget, $r_A = 1 \pm \Theta(\varepsilon)$. But then $r_B \leq a_B(r) + \varepsilon \leq 2\Theta(\varepsilon) + \varepsilon < 1 - \varepsilon$ for $\varepsilon \ll 1$. Contradiction.

- If $r_B < 1 - \varepsilon$, then $r_B \leq \beta + \varepsilon \leq K - \Theta(\frac{\varepsilon}{1-\beta})$, where the first inequality is by Proposition 1 and the second inequality holds for $\varepsilon \ll 1 - \beta$. Therefore, $r_u = 0 \pm \Theta(\varepsilon)$ and since $r_v \leq 1/4$ we have $r_a = 0 \pm \Theta(\varepsilon)$. This implies $a_B(r) \geq 2(1 - \varepsilon)(1 - \Theta(\varepsilon)) \geq 1 + \varepsilon$ for $\varepsilon \ll 1$. This implies $r_B \geq 1 - \varepsilon$. Contradiction.

---

[17]The threshold for $\varepsilon$ is $\Theta((1-\beta)^2)$. The fact that this is nonlinear in $1-\beta$ is not a problem. Recall that we did *not* assume that $\varepsilon = \Theta(1-\beta)$ if $\varepsilon \ll 1 - \beta$.

If $\alpha < \beta = 1$, let $e_B = c_{s,B}^A = 4/5$ and $K = \frac{3\alpha+1}{4}$ and $L = \frac{\alpha+3}{4}$, and keep everything else the same as above. Note that $\alpha < K < \frac{\alpha+1}{2} < L < 1$ evenly spaced and the output error of the cutoff gadget is $\Theta(\frac{\varepsilon}{1-\alpha})$. It is clear that property 1 follows just like above. We show that property 2 follows in a similar way to above. Assume that $r_a \leq 1/4$.

- If $r_B \geq 1 - \varepsilon$, then this implies $r_B \geq L + \Theta(\varepsilon)$ for $\varepsilon \ll 1 - \alpha$, so $r_A = 1 \pm \Theta(\varepsilon)$ and thus $r_B \leq a_B(r) + \varepsilon \leq 4/5 + 4/5 \cdot \Theta(\varepsilon) + \varepsilon < 1 - \varepsilon$ for $\varepsilon \ll 1$. Contradiction.

- If $r_B < 1 - \varepsilon$, we must have $a_B(r) < 1 + \varepsilon$ and $r_B = a'_B(r) \pm \varepsilon \leq a_B(r) - (1 - \alpha)e_B + \varepsilon < 1 - (1 - \alpha)e_B + 2\varepsilon$. The middle inequality is by definition of $a'_B$ in case $\beta = 1$. We further receive

$$
\begin{aligned}
1 - (1 - \alpha)e_B + 2\varepsilon &= 1 - \frac{4}{5}(1 - \alpha) + 2\varepsilon \\
&= \frac{3\alpha + 1}{4} - \frac{1}{20}(1 - \alpha) - 2\varepsilon \\
&= K - \frac{1}{20}(1 - \alpha) - 2\varepsilon \leq K - \Theta\left(\frac{\varepsilon}{1 - \alpha}\right)
\end{aligned}
$$

where the last line holds for $\varepsilon \ll 1 - \alpha$. This implies $r_A = 0 \pm \Theta(\varepsilon)$ and $a_B(r) \geq 4/5 + 4/5(1 - \varepsilon)(1 - \Theta(\varepsilon)) \geq 1 + \varepsilon$ for $\varepsilon \ll 1$ and this implies $r_B \geq 1 - \varepsilon$. Contradiction. $\square$

Theorem 1 now follows by application of the previous lemma to our financial Boolean circuit.

*Proof of Theorem 1, hardness.* Reduction from CIRCUIT SATISFIABILITY. Given a Boolean circuit $C$, apply Corollary 2 to construct the financial system $X$ with output bank $v$ corresponding to $C$. Apply the system from Lemma 5 to $X$ and $v$ (where by "application" we mean the same like for gadgets) to construct an extended system $X'$. Let $\varepsilon \ll 1$ and solve $\varepsilon$-HASCLEARING$^{\alpha,\beta}$ for $X'$.

If $C$ has a satisfiable assignment $\chi$, this yields an exact solution for $\chi$ where $r_v = 1 \geq 3/4$, so we can extend this to an exact solution of $X'$. If $C$ has no satisfying assignment, then any solution to $X$ satisfies $r_v = 0 \pm \Theta(\varepsilon)$ and therefore, no $\varepsilon$-solution for $X'$ exists. As these are the only two cases, any algorithm for $\varepsilon$-HASCLEARING$^{\alpha,\beta}$ must terminate on this instance and return YES if $C$ is satisfiable and NO otherwise. $\square$

# 4 The Complexity of the Search Problem without Default Costs

We will now focus on financial systems without default costs, i.e., where $\alpha = \beta = 1$. In these systems, we know that a solution always exists:

**Theorem** (Schuldenzucker, Seuken and Battiston, 2019, Theorem 2). *Any financial system $(N, e, c, \alpha = 1, \beta = 1)$ has an exact solution.*

The proof of the above theorem is by Kakutani's fixed point theorem and thus not constructive. In this section, we study the associated search problem. Since it may still be the case that all solutions are irrational (see Appendix A), we study the associated approximation problem of computing an $\varepsilon$-solution.

**Theorem 2.** *For $\varepsilon \ll 1$, the total search problem $\varepsilon$-FindClearing, defined as follows, is PPAD-complete: Given a non-degenerate financial system $X = (N, e, c, \alpha = 1, \beta = 1)$, compute an $\varepsilon$-solution.*

The theorem immediately implies that no polynomial-time approximation scheme (PTAS) exists, unless P=PPAD.

By Lemma 1, there is always an $\varepsilon$-solution of polynomial length. Recall from Proposition 1 that for $\alpha = \beta = 1$, an $\varepsilon$-solution is the same as an $\varepsilon$-approximate fixed point of the update function, i.e., an $r \in [0, 1]^N$ such that $F(r) = r \pm \varepsilon$. Note further that, since $\alpha = \beta = 1$ and we assume non-degeneracy, $F$ simplifies to:

$$F_i(r) = \begin{cases} 1 & \text{if } c_{i,j}^k = 0 \ \forall j \in N, \ k \in N \cup \{\emptyset\} \\ \left[\frac{a_i(r)}{l_i(r)}\right] & \text{otherwise.} \end{cases}$$

Membership in PPAD now easily follows.

*Proof of Theorem 2, membership.* The proof is similar to the proof of Lemma 1. By the above considerations, $F$ is polynomially continuous. And finding a (Brouwer) fixed point of a polynomially continuous function is in PPAD (Papadimitriou, 1994). $\square$

The remainder of this section is dedicated to showing PPAD-hardness via reduction from generalized circuits.

## 4.1 Generalized Circuits

A generalized circuit (Chen, Deng and Teng, 2009) consists of nodes interconnected by arithmetic or comparison gates. In contrast to regular arithmetic or Boolean circuits, generalized circuits may contain cycles, which turns finding a consistent assignment of node values into a non-trivial fixed point problem. Rubinstein (2018) introduced a variant of generalized circuits that is well-suited for our purposes. To make our

**Figure 6** Conditions that should hold at a gate $g$ for an $\varepsilon$-solution $x$ of a generalized circuit. Assume that the inputs of $g$ are called $a$ and $b$ (if any) and the output is called $v$. The gates $C_\zeta$, $C_{\times\zeta}$, and $C_{>\zeta}$ take an additional parameter $\zeta \in [0,1]$.

$$
\begin{aligned}
g = C_\zeta &\quad \Rightarrow \quad x[v] = \zeta \pm \varepsilon \\
g = C_+ &\quad \Rightarrow \quad x[v] = [x[a] + x[a]] \pm \varepsilon \\
g = C_- &\quad \Rightarrow \quad x[v] = [x[a] - x[b]] \pm \varepsilon \\
g = C_{\times\zeta} &\quad \Rightarrow \quad x[v] = \zeta \cdot x[a] \pm \varepsilon \\
g = C_{>\zeta} &\quad \Rightarrow \quad x[a] < \zeta - \varepsilon \Rightarrow x[v] = 0 \pm \varepsilon \\
&\quad\quad\quad\quad\, x[a] > \zeta + \varepsilon \Rightarrow x[v] = 1 \pm \varepsilon
\end{aligned}
$$

reduction to financial systems as simple as possible, we consider a reduced set of gates, which does not change the computational properties (see Appendix B for a detailed comparison).

**Definition 5** (Generalized Circuit and Approximate Solution)**.** A *generalized circuit* is a collection of *nodes* and *gates*, where each node is labeled *input* of any number of gates (including zero) and *output* of at most one gate. Inputs to the same gate are distinguishable from each other. Each gate has one of the following types: $C_\zeta$ (constant, no inputs), $C_{\times\zeta}$ (scaling, one input), $C_+$ or $C_-$ (addition and subtraction, two inputs), or $C_{>\zeta}$ (comparison to a constant, one input). For the gate types $C_\zeta$, $C_{\times\zeta}$, and $C_{>\zeta}$, a numeric parameter $\zeta \in [0,1]$ is specified in addition to the input and output nodes of the gate. The *length* of a generalized circuit is the number of bits needed to describe the circuit, including the nodes, the mapping from nodes to inputs and outputs of gates, and numeric parameters $\zeta$ involved.

For $\varepsilon \geq 0$, an *$\varepsilon$-solution* of a generalized circuit is a mapping $x$ that assigns to each node $v$ a value $x[v] \in [0,1]$ such that the constraints in Figure 6 hold at each gate of type $g$ with inputs $a$ and $b$ (if any) and output $v$.

We know from prior work that finding an $\varepsilon$-solution is hard:

**Theorem** (Essentially Rubinstein (2018))**.** *For a (constant) sufficiently small $\varepsilon$, the total search problem $\varepsilon$-GCircuit, defined as follows, is PPAD-complete: Given a generalized circuit, find an $\varepsilon$-solution.*

## 4.2 Reducing Generalized Circuits to Financial Systems

We now show how to encode a generalized circuit into a financial system via financial system gadgets corresponding to the five gate types. Any $\varepsilon$-solution to the financial system will give rise to a $\Theta(\varepsilon)$-solution of the generalized circuit. Compared to Section 3, our gadgets need to be more precise because we do not only have to map between the appropriate parts of the set $[0, 1/4] \,\dot\cup\, [3/4, 1]$ to each other, but we have to encode arithmetic operations on continuous inputs in $[0, 1]$. The fact that we do not have default costs in this section will help us achieve this higher precision. In our gadgets, all banks except for the source and sink banks will have liabilities constant 1. We therefore have $r_i = F_i(r) \pm \varepsilon = [a_i(r)] \pm \varepsilon$ in any $\varepsilon$-solution.

**Figure 7** Constant Gadget

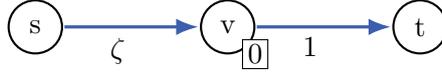

**Figure 8** Inverter Gadget

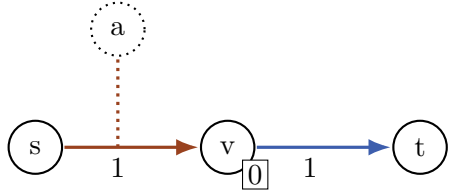

Our simplest gadget establishes a constant recovery rate at the output bank:

**Lemma 6** (Constant Gadget). *Let $\zeta \in [0, 1]$. If $\alpha = \beta = 1$, there is a financial system gadget with no input banks that implements the property $r_v = \zeta \pm \Theta(\varepsilon)$.*

*Proof.* Consider Figure 7. Clearly, $a_v(r) = \zeta(1 \pm \varepsilon) = \zeta \pm \varepsilon$ and $r_v = [a_v(r)] \pm \varepsilon = \zeta \pm 2\varepsilon$. □

An important building block for the following constructions is a gadget that "inverts" the recovery rate of a bank.

**Lemma 7** (Inverter Gadget). *If $\alpha = \beta = 1$, there is a financial system gadget with one input bank that implements the property $r_v = 1 - r_a \pm \Theta(\varepsilon)$.*

*Proof.* Consider Figure 8. Clearly, $a_v(r) = (1 \pm \varepsilon)(1 - r_a) = 1 - r_a \pm \varepsilon$ and $r_v = [a_v(r)] \pm \varepsilon = [1 - r_a] \pm 2\varepsilon = 1 - r_a \pm 2\varepsilon$. □

Note that we could *not* have used the inverter gadget as a Boolean NOT gadget in Section 3 because i) it relies on the assumption $\alpha = \beta = 1$ and ii) it accumulates errors, i.e., $2n$ inverters in a row yield $r_a \pm \Theta(n\varepsilon)$, not $r_a \pm \varepsilon$. We proceed with the addition and subtraction gadgets, which are slight variants of each other.

**Lemma 8** (Sum Gadget). *If $\alpha = \beta = 1$, there is a financial system gadget with two input banks that implements the property $r_v = [r_a + r_b] \pm \Theta(\varepsilon)$.*

**Figure 9** Sum Gadget. $\bar{a}$ and $\bar{b}$ are the outputs of inverters applied to $a$ and $b$, respectively.

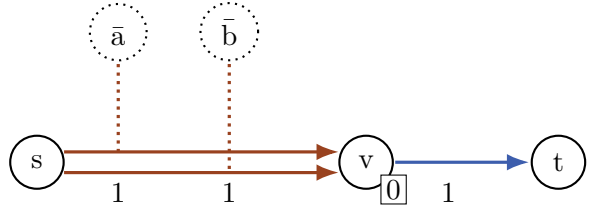

*Proof.* Apply inverter gadgets (Lemma 7) to both $a$ and $b$ and call the output banks $\bar{a}$ and $\bar{b}$, respectively. Now apply the gadget in Figure 9. Then:

$$r_v = [(1 \pm \varepsilon)(1 - r_{\bar{a}}) + (1 \pm \varepsilon)(1 - r_{\bar{b}})] \pm \varepsilon = [r_a + r_b] \pm 3\varepsilon \qquad \square$$

Note the similarity of Figure 9 with Figure 4, which was used in the construction of the NAND gadget in Section 3. Indeed. a similar operation is performed given that $a \text{ NAND } b = \neg a \vee \neg b$ and addition is somewhat similar to Boolean "$\vee$". Note however that these two constructions need to deal with different challenges: the NAND gadget needs to work with default costs and the sum gadget needs to provide a correct sum across all input values, not just approximate Boolean values.

**Lemma 9** (Difference Gadget). *There is a financial system gadget without default costs with two input banks that implements the property $r_v = [r_a - r_b] \pm \Theta(\varepsilon)$.*

*Proof.* Apply an inverter gadget (Lemma 7) to $a$ and call the output bank $\bar{a}$. Apply the gadget in Figure 9 to $\bar{a}$ and $\bar{b} := b$ and call the output bank $u$. From the proof of the previous lemma we know that $r_u = [1 - r_a + r_b] \pm \Theta(\varepsilon)$. Now apply an inverter to $u$ and call the output bank $v$. It follows that

$$r_v = 1 - [1 - r_a + r_b] \pm \Theta(\varepsilon) = [r_a - r_b] \pm \Theta(\varepsilon),$$

where the last equality follows by case distinction. $\qquad \square$

Our last two gadgets are scaling and comparison. Scaling is easily achieved by noting that a chain of two inverters approximately copies the input. We then adjust the notional in one of those gadgets. This happens to be a degenerate variant of the cutoff gadget.

**Lemma 10** (Scaling Gadget). *Let $\zeta \in [0, 1]$. If $\alpha = \beta = 1$, there exists a financial system gadget that implements the property $r_v = \zeta r_a \pm \Theta(\varepsilon)$.*

*Proof.* We use the financial system in Figure 3 from Section 3 with $\gamma = 1$ and $\delta = \zeta$. We have $r_u = 1 - r_a \pm \Theta(\varepsilon)$ like in the inverter gadget and thus $a_v(r) = \zeta(1 \pm \varepsilon)(1 - $ $(1 - r_a \pm \Theta(\varepsilon))) = \zeta(1 \pm \varepsilon)(r_a \pm \Theta(\varepsilon)) = \zeta r_a \pm \Theta(\varepsilon)$. And $r_v = [a_v(r)] \pm \varepsilon = \zeta r_a \pm \Theta(\varepsilon)$. Here we use $\zeta, r_a \leq 1$ to bound the error. $\qquad \square$

For the comparison gate, the cutoff gadget (Lemma 3) almost does what we want. However, the comparison gate makes demands to low brittleness (at the order of $\varepsilon$) and low output error (*also* at the order of $\varepsilon$) that the cutoff gadget is not able to achieve. Fortunately, we can use our previously introduced reset gadget (which itself happens to be another incarnation of the cutoff gadget) to fix the output error.

**Lemma 11** (Comparison Gadget). *Let $\zeta \in [0, 1]$. If $\alpha = \beta = 1$, there exists a financial system gadget with one input bank that implements the following property:*

$$r_a \leq \zeta - \Theta(\varepsilon) \Rightarrow r_v = 0 \pm \Theta(\varepsilon)$$
$$r_a \geq \zeta + \Theta(\varepsilon) \Rightarrow r_v = 1 \pm \Theta(\varepsilon)$$

*Proof.* Let $C > 0$ be a constant such that $C$ is an upper bound on the implicit constant factors in the expressions $\Theta(\frac{\varepsilon}{L-K})$ on the right-hand sides of Lemma 3 (i.e., the output error; we can choose $C = 3$). Let $c = 2C$. Assume WLOG that $c\varepsilon < \zeta < 1 - c\varepsilon$. If this is not the case, we can simply give $v$ recovery rate constant 0 or 1. Apply now a cutoff gadget (Lemma 3) with $K = \zeta - c\varepsilon$ and $L = \zeta + c\varepsilon$ and call the output gate $u$. Then apply a reset gadget (Corollary 1) to $u$ and call the output $v$.

By the cutoff gadget, if $r_a \leq K - \Theta(\varepsilon) = \zeta - \Theta(\varepsilon)$, then $r_u = 0 \pm C\frac{\varepsilon}{L-K} = 0 \pm C\frac{\varepsilon}{2c\varepsilon} = 0 \pm 1/4$. Thus, by the reset gadget, $r_v = 0 \pm \Theta(\varepsilon)$. Likewise for $r_a \geq \zeta + \Theta(\varepsilon)$. $\qquad \square$

With all gadgets in place, we can connect our gadgets to represent a generalized circuit and prove PPAD-hardness.

*Proof of Theorem 2, hardness.* Reduction from $\Theta(\varepsilon)$-GCIRCUIT. Let $C$ be a generalized circuit. We construct a financial system. For each node $v$ of $C$, add a bank with no assets or liabilities and identify that bank with $v$. For each gate $g$ of $C$, execute the corresponding gadget from this section with the appropriate input banks and identify the output bank of the gadget with $g$. Finally, if a node $v$ is the output of gate $g$, take a copy of the scaling gadget with $\zeta = 1$ (this is the same as two inverters connected) and identify the input of the gadget with $g$ and the output of the gadget with $v$.[18]

---

[18] In the language of gadgets, this operation can be interpreted as applying the constructed financial system, say $X_0$, and the $\times 1$-scaling gadget "to each other." Note that in $X_0$, $v$ has no assets and liabilities and can thus be considered an input bank. The result of the "application" will be an extension of $X_0$ at all banks but $v$ and of the scaling gadget at all banks but its input. By Remark 2., this implies that all gadgets still behave as expected. In particular, we have $r_v = r_g \pm \Theta(\varepsilon)$ in any $\varepsilon$-solution.

Now, by the gadgets, if $\varepsilon \ll 1$ and $r$ is an $\varepsilon$-solution for the financial system, setting $x[v] = r_v$ for all nodes yields a $\Theta(\varepsilon)$-solution of the generalized circuit. And finding this is hard for $\varepsilon \ll 1$. $\qquad \square$

# 5   The Complexity of Determining Defaults

Given the results in the previous two sections, one may wonder if we can somehow pin down the "origin" of the computational complexity. What is it really that CDSs *do* to a financial system that makes the decision and search problems so much harder to solve? In this section and the next we explore this question.

To understand where the computational complexity comes from, we look for ways to circumvent it. One way to do this might be to ask for less information than the recovery rates themselves. If it is easy to find some bounds on the recovery rates, for example, this could already be very useful. The minimum level of detail we will likely be interested in is which banks default. We define a *default set* as a collection of banks that (approximately) default in some (approximate) solution. A default set thus provides a kind of "coarse representation" of a solution to the clearing problem.

**Definition 6.** Let $X = (N, e, c, \alpha, \beta)$ be a financial system and let $\varepsilon \geq 0$. A set $D \subseteq N$ is called an *$\varepsilon$-default set* for $X$ if there is an $\varepsilon$-solution $r$ for $X$ such that for all $i \in N$:

$$i \notin D \Rightarrow a_i(r) \geq (1 - \varepsilon) l_i(r)$$
$$i \in D \Rightarrow a_i(r) < (1 + \varepsilon) l_i(r).$$

In this case, we call $r$ and $D$ *$\varepsilon$-compatible*.

We have relaxed the notion of being in default in the same way as in the definition of an $\varepsilon$-solution. Again, this ensures that the problem will not be hard for the wrong reasons, namely because of "knife-edge" defaults, where a small error in the assets or liabilities could otherwise determine whether a bank defaults and thus lead to a large error in the recovery rate. Banks at the edge of default can instead be considered either of the two. A side effect of this liberty is that, say, $i \notin D$ does *not* generally imply $r_i = 1 \pm \varepsilon$. It could also be that $a_i(r) \in [(1 - \varepsilon) l_i(r),\ (1 + \varepsilon) l_i(r))$ and $r_i = \frac{a_i'(r)}{l_i(r)}$. That is, the decision about default in $D$ and $r$ need not coincide.

In the remainder of this section, we show that computational complexity does not arise exclusively from the need to compute precise numeric values for the recovery rates. Rather, computational problems are already hard at the level of default sets. We study a decision and a search variant.

### 5.1 Deciding the Default of a Given Bank

In the decision variant of our problem, we ask if there is a solution where a specified bank defaults. This is a basic question a regulator might ask in a stress test: given a certain shock scenario, will this make it necessary to save AIG *again*? If there are multiple solutions and AIG only defaults in some of them, our answer should still be "Yes," i.e., we should consider the worst case.

We can consider this decision problem for any value of the default cost parameters. If there are default costs, a naive framing of the problem would contain the question if there is any solution in the first place. Since this is not what we are interested in at this point, we exclude it by a promise. We then consider a distinction variant similar to $\varepsilon$-HasClearing$^{\alpha,\beta}$ (Theorem 1). The proof is rather straightforward using our financial Boolean circuits from Section 3. The key technical step is to notice that assets are either very small or very large compared to liabilities, so that defaults are never ambiguous.

**Theorem 3.** *For any fixed $\alpha, \beta \in [0, 1]$ and any $\varepsilon \ll 1$, the following promise problem is NP-complete: Given a non-degenerate financial system $X = (N, e, c, \alpha, \beta)$, and a bank $i \in N \ldots$*

- *if $X$ has an exact solution and there is an exact default set $D$ such that $i \in D$, return YES.*

- *if $X$ has an exact solution and there is no $\varepsilon$-default set $D$ such that $i \in D$, return NO.*

*Proof.* Membership: By the proof of Lemma 1, if $r$ is an exact solution and $D$ exactly compatible, we receive a polynomial-length $\varepsilon$-solution with which $D$ is $\varepsilon$-compatible with via rounding. Hence, it is enough to check all $\varepsilon$-solutions of a certain polynomial maximum length. Note that if $r$ is an $\varepsilon$-solution, then $i \in D$ for some $\varepsilon$-compatible default set $D$ iff $a_i(r) < (1 + \varepsilon)l_i(r)$.

Hardness: Reduction from CIRCUIT FALSIFIABILITY. Given a Boolean circuit $C$, consider the financial Boolean circuit system $X$ from Corollary 2 and the output node $v =: i$. Recall that $v$ is the output of a NAND gadget (Lemma 4). If $C$ is falsifiable, let $r$ be the exact solution corresponding to a falsifying assignment and $D$ its exactly compatible default set. By the proof of Lemma 4, we then have $a_v(r) = 0 < 1 = l_i(r)$ and thus $v \in D$. If $C$ is not falsifiable, let $\varepsilon \ll 1$, let $r$ be any $\varepsilon$-solution and $D$ $\varepsilon$-compatible. By Corollary 2, the inputs to $v$'s NAND gadget are approximately Boolean and not both TRUE and again by the proof of the lemma, we have $a_v(r) \geq 1 + \varepsilon$, so $v \notin D$. □

The theorem immediately implies that the following problem is NP-hard: Given a non-degenerate financial system with the promise that it has an exact solution and a bank $i$, decide if $i \in D$ for some $\varepsilon$-default set $D$. The problem may not be in NP because it is not guaranteed that every $\varepsilon$-solution has polynomial length. See the discussion after Theorem 1. Note in particular that it is not clear how to check if a given $D \subseteq N$ is an $\varepsilon$-default set.

It is easy to see that the distinction problem is still NP-complete if we replace "$\in D$" by "$\notin D$" (proof by negation of the circuit) and it is coNP-complete if we replace "there exists $D$" by "for all $D$" (proof by reduction from Circuit Contradiction).

## 5.2 Finding Default Sets Without Default Costs

We return to the search problem without default costs. It is trivial to, given an $\varepsilon$-solution, compute a default set compatible with it, while the converse is not so clear. Thus, finding $\varepsilon$-default sets may a priori be easier than finding the $\varepsilon$-solution itself. We will show that this is not the case.

**Theorem 4.** *For $\varepsilon \ll 1$, the following problem is PPAD-complete: Given a non-degenerate financial system $X = (N, e, c, \alpha = 1, \alpha = 1)$, compute an $\varepsilon$-default set.*

In some domains, statements like this follow trivially hardness of the respective continuous variant. For example, in a two-player normal-form game, the supports of the strategies of the two players (i.e., the strategies that are played with nonzero probability) could be taken as a "coarse representation" of a Nash equilibrium, similar to default sets in our case. Finding the supports of a Nash equilibrium is trivially PPAD-complete because a Nash equilibrium can be reconstructed from its supports via linear programming, and finding Nash equilibria in two-player games is hard (Chen, Deng and Teng, 2009).[19] Given an $\varepsilon$-default set however, there does not seem to be an easy way to reconstruct a corresponding $\varepsilon$-solution, given that already the assets $a_i$ can contain terms like $r_j(1 - r_k)$, which are non-linear, non-convex/concave, and non-monotonic. This remains true for the particular construction we perform in Section 4. Here, the assets of the relevant banks contain terms of form $r_s(1 - r_a)$ where $s$ is the source bank and $r_s \in [1 - \varepsilon, 1]$. It may be tempting to just assume $r_s = 1$ to make the problem linear. However, given a default set of an $\varepsilon$-solution where (say) $r_s = 1 - \varepsilon$, assuming $r_s = 1$ implicitly introduces an error of $\varepsilon$ into $r_s$. As the notionals in the system can be at the order of $1/\varepsilon$,[20] this may in turn imply a large (constant in $\varepsilon$) change in the recovery rate of some other bank. No $\varepsilon$-solution with such a recovery rate may exist.

Rather than reconstructing an $\varepsilon$-solution to the financial system from a default set, we introduce a new discrete variant of the $\varepsilon$-GCircuit problem that is still PPAD-complete and that may be of independent interest. We then show that any $\varepsilon$-default set gives rise to a solution for the discrete $\Theta(\varepsilon)$-GCircuit problem.

---

[19] $\varepsilon$ must be polynomial, not constant, in the size of the game in this particular example.

[20] The comparison gadget (Lemma 11) introduces notionals of order $1/\varepsilon$.

**Figure 10** Constraints that need to hold at each gate $g$ when $x$ is an $\varepsilon$-solution for a generalized circuit compatible with an assignment $d$. Let $a$ and $b$ be the inputs (if any) and $v$ the output of $g$.

$$
\begin{aligned}
g = C_+ \ &\Rightarrow\ d_g \in \{L, M\} \ &&\Rightarrow\ x[a] + x[b] \leq 1 + \varepsilon \wedge x[v] = x[a] + x[b] \pm \varepsilon \\
&\phantom{\Rightarrow\ } d_g = H \ &&\Rightarrow\ x[a] + x[b] \geq 1 - \varepsilon \wedge x[v] = 1 \pm \varepsilon \\
g = C_- \ &\Rightarrow\ d_g = L \ &&\Rightarrow\ x[a] - x[b] \leq \varepsilon \qquad \wedge x[v] = 0 \pm \varepsilon \\
&\phantom{\Rightarrow\ } d_g \in \{M, H\} \ &&\Rightarrow\ x[a] - x[b] \geq -\varepsilon \quad \wedge x[v] = x[a] - x[b] \pm \varepsilon \\
g = C_{>\zeta} \ &\Rightarrow\ d_g = L \ &&\Rightarrow\ x[a] \leq \zeta + \varepsilon \qquad \wedge x[v] = 0 \pm \varepsilon \\
&\phantom{\Rightarrow\ } d_g = M \ &&\Rightarrow\ x[a] = \zeta \pm \varepsilon \\
&\phantom{\Rightarrow\ } d_g = H \ &&\Rightarrow\ x[a] \geq \zeta - \varepsilon \qquad \wedge x[v] = 1 \pm \varepsilon
\end{aligned}
$$

### 5.3 The discrete $\varepsilon$-GCircuit problem

The main idea for our discrete version of $\varepsilon$-GCircuit is that the constraints for a generalized circuit (Figure 6) are piecewise linear with a finite number of "cutoff" or "decision points." For example, the $C_+$ gate corresponds to either a sum or a constant depending on whether the sum of its inputs lies in $[0, 1]$ or $(1, \infty)$. These "decision points," defined an appropriate way, allow us to reconstruct a solution by solving a linear feasibility problem. Our approach is similar in spirit to Vazirani and Yannakakis (2011), who split equilibrium computation in Fisher markets into two steps: first, a PPAD-complete problem is solved to determine the combinatorial structure of the equilibrium. Second, the exact numbers are computed in polynomial time. We relax the definitions in a way that mirrors the definition of an $\varepsilon$-default set, allowing ambiguity when we are close to the respective "decision point," which leaves us with three gate types.[21] The $C_\zeta$ and $C_{\times\zeta}$ gates have linear constraints and therefore do not need any "decision" specified.

**Definition 7.** Let $C$ be a generalized circuit and $\varepsilon \geq 0$. A *discrete $\varepsilon$-solution* for $C$ assigns to each gate $g$ of $C$ a value $d_g \in \{H, M, L\}$ such that there exists an $\varepsilon$-solution $x$ for $C$ such that the constraints in Figure 10 hold for each gate $g$ with inputs $a$ and $b$ and output $v$. In this case, we call $x$ and $d$ $\varepsilon$-*compatible*.

Note that our definition is *monotonic* in $\varepsilon$: if $d$ is a discrete $\varepsilon$-solution and $\varepsilon' > \varepsilon$, then $d$ is also a discrete $\varepsilon'$-solution. This is what we typically expect of an approximate solution concept and we take it as an indication that our concept of a discrete $\varepsilon$-solution is natural. We will exploit monotonicity in our proofs below.

*Remark* 3. If $x$ is an $\varepsilon$-solution for $C$, we can define a discrete $\varepsilon$-solution $d_g$ as follows.

---

[21]This relaxation is crucial for our following reduction from $\varepsilon$-default sets. This is not just because $\varepsilon$-default sets have this kind of relaxation, but due to $\varepsilon$ errors in the solution itself. That is, even if we required exact compatibility in the definition of an $\varepsilon$-default set, we would have to make this relaxation here.

- If $g = C_+$, let $d_g = H$ if $x[a] + x[b] \geq 1$ and $d_g = M$ otherwise.

- If $g = C_-$, let $d_g = L$ if $x[a] - x[b] \leq 0$ and $d_g = M$ otherwise.

- If $g = C_{>\zeta}$, let

$$
d_g = \begin{cases} L & \text{if } x[a] < \zeta - \varepsilon \\ M & \text{if } x[a] = \zeta \pm \varepsilon \\ H & \text{if } x[a] > \zeta + \varepsilon. \end{cases}
$$

The fact that $x$ is an $\varepsilon$-solution will now ensure that the outputs match, too, and thus $d$ is a discrete $\varepsilon$-solution compatible with $x$.

**Theorem 5.** *For a (constant) $\varepsilon \ll 1$, the following total search problem, which we call the* discrete $\varepsilon$-GCircuit *problem, is PPAD-complete: Given a generalized circuit, find a discrete $\varepsilon$-solution.*

*Proof.* Membership is obvious by reduction to the (continuous) $\varepsilon$-GCircuit problem. For hardness, we perform reduction *from* the continuous problem. Let $C$ be a generalized circuit and let $d$ be a discrete $\varepsilon$-solution for $C$. Consider the linear feasibility problem (LFP) with a variable $x[v]$ for each node $v$ of $C$ and the following constraints:

1. For each $v$, add the constraint $0 \leq x[v] \leq 1$.

2. For each gate $g$ of type $C_+$, $C_-$, or $C_{>\zeta}$, add the constraint from Figure 6 corresponding to the value of $d_g$.

3. For each gate $g$ of type $C_\zeta$ or $C_{\times\zeta}$, add the constraint from Figure 6 corresponding to the type of $g$.

Note that all constraints are linear. It is easy to verify that a vector $x$ is feasible for the LFP iff it is an $\varepsilon$-solution $\varepsilon$-compatible with $d$. By assumption, such an $x$ exists and we can find it in polynomial time via the LFP. $\qquad \square$

The proof of the theorem implies that a generalized circuit always has an exact solution of polynomial length. This was previously noted by Etessami and Yannakakis (2010) in their study of their FIXP complexity class. To see it from our proof, let $d$ be the default set of an exact solution (which exists by Kakutani's fixed-point theorem) and solve the LFP for $\varepsilon = 0$.

### 5.4  $\varepsilon$-Default Sets Inform a Discrete $\Theta(\varepsilon)$-Solution

We now re-examine the sum, difference, and comparison gadget from Section 4. We show that the default states of some banks can be used to define a discrete $\Theta(\varepsilon)$-solution for the respective gate type that will be $\Theta(\varepsilon)$-compatible with any $\varepsilon$-solution of the gadget.

*Remark* 4 (Bounding Recovery Rates via Default Sets). In contrast to the case with default costs, if $\alpha = \beta = 1$, we can bound the recovery rates based on the default set. More in detail, if $r$ is an $\varepsilon$-solution $\varepsilon$-compatible with $D$, then:

$$i \notin D \Rightarrow r_i = 1 \pm 2\varepsilon$$

$$i \in D \Rightarrow r_i = \frac{a_i(r)}{l_i(r)} \pm 2\varepsilon$$

This follows by case distinction using the equivalence between an $\varepsilon$-solution and an $\varepsilon$-approximate fixed point of $F$.

If $G$ is a gadget with input banks $A$ and $P : 2^N \times [0,1]^N \rightarrow \{\text{True}, \text{False}\}$ is a property, we say that $G$ *implements* $P$ *on default sets* if for $\varepsilon \ll 1$, any $\varepsilon$-solution $r$ on $N \setminus A$, and any $D \subseteq N$ $\varepsilon$-compatible with $r$, $P(D, r)$ holds.

For the sum gadget, we can simply consider the default state of the output bank.

**Lemma 12** (Default Set of the Sum Gadget). *If $\alpha = \beta = 1$, then the sum gadget (Lemma 8) implements the following property on default sets:*

$$v \notin D \Rightarrow r_a + r_b \geq 1 - \Theta(\varepsilon) \wedge r_v = 1 \pm \Theta(\varepsilon)$$

$$v \in D \Rightarrow r_a + r_b < 1 + \Theta(\varepsilon) \wedge r_v = r_a + r_b \pm \Theta(\varepsilon)$$

*Proof.* Recall that $a_v(r) = r_a + r_b \pm \Theta(\varepsilon)$ and $l_v(r) = 1$. The statement now follows from Remark 4 and the fact that $r_v = [r_a + r_b] \pm \Theta(\varepsilon)$. ☐

For the difference gadget, the default state of the output bank is not informative. To see this, recall that this bank is the output of an inverter gadget. In the inverter gadget, though, the output bank is never unambiguously not in default because its assets are at most 1 (see Figure 8). Thus, we could have $v \in D$ independently of the input or output recovery rates. To extract information from the default set, we need to consider the *input* bank to the inverter gadget instead.

**Lemma 13** (Default Set of the Difference Gadget). *Let $\alpha = \beta = 1$ and consider the difference gadget (Lemma 9). Let $u$ be the intermediate bank in that gadget. Then the gadget implements the following property on default sets:*

$$u \notin D \Rightarrow r_a - r_b \leq \Theta(\varepsilon) \quad \wedge r_v = 0 \pm \Theta(\varepsilon)$$

$$u \in D \Rightarrow r_a - r_b > -\Theta(\varepsilon) \wedge r_v = r_a - r_b \pm \Theta(\varepsilon)$$

*Proof.* Recall that we have $a_u(r) = 1 - r_a + r_b \pm \Theta(\varepsilon)$ and $r_v = 1 - r_u \pm \Theta(\varepsilon)$. The statement now follows just like in Lemma 12. ☐

For the comparison gadget, we proceed in a similar way and consider the default states in the first cutoff gadget (before the reset gadget). We need to consider the default states of both of the banks in this gadget to determine in which of the three possible states the gadget is. Compared to the other two gadgets, we need to consider the details of the construction in much greater detail.

**Lemma 14** (Default Set of the Comparison Gadget)**.** *Let $\alpha = \beta = 1$ and consider the comparison gadget (Lemma 11). Let $u_1$ and $v_1$ be the intermediate banks that correspond to the first cutoff gadget. Then the comparison gadget implements the following property on default sets:*

$$u_1 \notin D \wedge v_1 \in D \Rightarrow r_a \leq \zeta + \Theta(\varepsilon) \wedge r_v = 0 \pm \Theta(\varepsilon)$$
$$u_1 \in D \wedge v_1 \notin D \Rightarrow r_a \geq \zeta - \Theta(\varepsilon) \wedge r_v = 1 \pm \Theta(\varepsilon)$$
$$u_1 \in D \wedge v_1 \in D \Rightarrow r_a = \zeta \pm \Theta(\varepsilon)$$
$$u_1 \notin D \wedge v_1 \notin D \text{ is impossible.}$$

*Proof.* Recall from the proof of Lemma 11 that the first cutoff gadget has parameters $K = \zeta - c\varepsilon$ and $L = \zeta + c\varepsilon$ where $c > 4$ is a sufficiently large constant. Recall from the definition of the cutoff gadget (Lemma 3) that this implies for the notionals in Figure 3 that

$$\gamma = \frac{1}{1-K} = \frac{1}{1-\zeta+c\varepsilon}$$
$$\delta = \frac{1-K}{L-K} = \frac{1-\zeta+c\varepsilon}{2c\varepsilon} = \frac{1-\zeta}{2c\varepsilon} + \frac{1}{2}.$$

Assume first that $u_1 \notin D$. By definition of an $\varepsilon$-default set and $l_{u_1}(r) = 1$, we have $1 - \varepsilon \leq a_{u_1}(r) = \gamma(1 - r_a)$. Rearranging yields $r_a \leq \zeta - c\varepsilon + \varepsilon(1 - \zeta + c\varepsilon) \leq \zeta - (c-2)\varepsilon$ if $\varepsilon \ll c$. We further must have $r_{u_1} \geq 1 - 2\varepsilon$, so $r_{v_1} \leq a_{v_1}(r) + \varepsilon \leq \delta \cdot 2\varepsilon + \varepsilon = (1-\zeta)/c + 2\varepsilon \leq 1/4$ for $\varepsilon \ll 1$. This implies $v_1 \in D$ and, as $v_1$ is input to a reset gadget with output $v$, $r_v = 0 \pm \Theta(\varepsilon)$.

Assume next that $u_1 \in D$. Then by Remark 4, $r_{u_1} = a_{u_1}(r) \pm 2\varepsilon = \gamma(1 - r_a) \pm (\gamma + 2)\varepsilon$.

If now $u_1 \in D \wedge v_1 \notin D$, then

$$\begin{aligned}
1 - \varepsilon \leq a_{v_1}(r) &\leq \delta(1 - r_{u_1}) \\
&\leq \delta(1 - (\gamma(1 - r_a) - (\gamma + 2)\varepsilon)) \\
&= \delta - \delta\gamma + \delta\gamma r_a + (\delta\gamma + 2\delta)\varepsilon.
\end{aligned}$$

Rearranging yields: $r_a \geq \frac{1}{\delta\gamma} - \frac{1}{\gamma} + 1 - \left(1 + \frac{2}{\gamma} + \frac{1}{\delta\gamma}\right)\varepsilon \geq L - 4\varepsilon = \zeta + (c-4)\varepsilon$, where the middle inequality follows using the identities $\frac{1}{\delta\gamma} = L - K$ and $\frac{1}{\gamma} = 1 - K$. Of course, $r_{v_1} = 1 \pm 2\varepsilon$ because $v_1 \notin D$ and thus $r_v = 1 \pm \varepsilon$.

If $u_1 \in D \wedge v_1 \in D$, then

$$
\begin{aligned}
1 + \varepsilon \geq a_{v_1}(r) &\geq \delta(1 - \varepsilon)(1 - r_{u_1}) \\
&\geq \delta(1 - (\gamma(1 - r_a) + (\gamma + 2)\varepsilon) - \delta\varepsilon \\
&= \delta - \delta\gamma + \delta\gamma r_a - (\delta\gamma + 3\delta)\varepsilon.
\end{aligned}
$$

Rearranging like above yields: $r_a \leq L + 5\varepsilon = \zeta + (c + 5)\varepsilon$. This bounds $r_a$ from above. To bound $r_a$ from below, notice that, since $u_1 \in D$, we have $1 + \varepsilon > a_{u_1}(r) \geq \gamma(1 - \varepsilon)(1 - r_a)$. This implies $r_a \geq 1 - \frac{1}{\gamma} \cdot \frac{1 + \varepsilon}{1 - \varepsilon} \geq 1 - \frac{1}{\gamma}(1 + 3\varepsilon) \geq K - 3\varepsilon \geq \zeta - (c - 2)\varepsilon$, where the second inequality holds for $\varepsilon \ll 1$. □

Note that the implicit constants in the $\Theta(\varepsilon)$ expressions in the above lemma are not the same. That is why the different cases in the previous lemma overlap.

Note that the proof of the previous lemma actually gives us slightly stronger bounds on $r_a$ than stated. For example, in case $u_1 \notin D \wedge v_1 \in D$, we receive from the proof that $r_a \leq \zeta - (c - 2)\varepsilon = \zeta - \Theta(\varepsilon)$, not just $r_a \geq \zeta + \Theta(\varepsilon)$. Beyond uniformity with the definition of a discrete $\varepsilon$-solution, the weaker version of the conditions has the benefit of monotonicity: the conditions continue to hold if we increase $\varepsilon$. We will exploit this in the following proof.

Using the above three lemmas, we can define a discrete $\varepsilon$-solution to a generalized circuit given an $\varepsilon$-default set. This proves our theorem:

*Proof of Theorem 4.* Let $C$ be a generalized circuit and let $X$ be the financial system without default costs corresponding to $C$ like in the proof of Theorem 2. Let $\varepsilon \ll 1$ and let $D$ be an $\varepsilon$-default set of $X$. We show that for some $\varepsilon' = \Theta(\varepsilon)$, $D$ induces a discrete $\varepsilon'$-solution $d$ of $C$. This proves the theorem because finding the latter is hard for $\varepsilon' \ll 1$. We define $d$ following the preceding lemmas. For each gate $g$ of $C$, . . .

- If $g = C_+$, consider the corresponding sum gadget and let $d_g = H$ if $v \in D$ and $d_g = M$ if $v \notin D$.

- If $g = C_-$, consider the corresponding difference gadget and let $d_g = L$ if $u \notin D$ and $d_g = M$ if $u \in D$.

- If $g = C_>$, consider the corresponding comparison gadget and let

$$
d_g = \begin{cases} L & \text{if } u_1 \notin D \wedge v_1 \in D \\ H & \text{if } u_1 \in D \wedge v_1 \notin D \\ M & \text{if } u_1 \in D \wedge v_1 \in D. \end{cases}
$$

Let $r$ be an $\varepsilon$-solution of $X$ $\varepsilon$-compatible with $D$ and let $\varepsilon' = \Theta(\varepsilon)$ be the maximum of all the incarnations of $\Theta(\varepsilon)$ in Lemma 12–14, in the proof of Theorem 2, and $\varepsilon$ itself. By the proof of Theorem 2, $r$ induces an $\varepsilon'$-solution of $C$ by restriction to the output banks of the gadgets corresponding to gates. Since the above lemmas still hold if one replaces every instance of $\Theta(\varepsilon)$ by $\varepsilon'$ (due to monotonicity), they imply that this induced $\varepsilon'$-solution of $C$ is $\varepsilon'$-compatible with $d$. Thus, $d$ is a discrete $\varepsilon'$-solution of $C$. $\hfill\square$

In this section, we have shown that already finding an $\varepsilon$-default set of a financial system with CDSs is hard. En-route, we have developed a general methodology to show that "coarse" or "discrete" versions of PPAD-hard search problems are hard. We believe that our methodology can be applied to other problems to receive this type of result when the reduction is sufficiently faithful to the structure of a generalized circuit. For example, Daskalakis, Goldberg and Papadimitriou (2009) introduced gadgets that encode a generalized circuit in a binary, degree-3 graphical game. It is not hard to show that the supports of certain players in these gadgets inform a discrete $\varepsilon$-solution of the generalized circuit. Thus, already finding the supports of an $\varepsilon$-Nash equilibrium in such a game is hard.[22] Unlike for two-player, $n$-action games, this result is not trivial because graphical games can contain nonlinear interactions and two-player games are not an immediate special case of graphical games. However, in this particular instance, the result can be shown more directly. This is because the game gadgets can easily be modified to ensure that players' utilities are linear combinations of other players' strategies (Daskalakis, Goldberg and Papadimitriou, 2009, Section 6.1) and then an equilibrium can be reconstructed from the supports using linear programming. Future work may well encounter other domains where, like in financial networks with CDSs, no such modification is possible and where our methodology can be of use.

# 6 Structural Restrictions

We continue our quest towards the "origin" of the computational complexity in financial networks with CDSs. In this section, we study under which restrictions on the network structure the distinction and search problems are still hard. This is important, following our original program of study, to understand how the informal "complexity" due to CDSs arises and to inform potential regulatory policies that reduce it.

## 6.1 Counterparty Risk and Fundamental Risk

Inspection reveals that that our gadgets, and thus all hard instances constructed in Section 3 and 4, share three properties that make them particularly simple financial systems:

---

[22]Daskalakis, Goldberg and Papadimitriou (2009) showed hardness when $\varepsilon$ decreases with the size of the game exponentially. Rubinstein (2018) extended their result to a constant $\varepsilon$ using the same gadgets. From this, we receive hardness of finding supports for constant $\varepsilon$ as well.

1. *Acyclic Liabilities:* The *liability graph,* where each writer of a contract is connected to the respective holder, is acyclic. In fact, this graph is a disjoint union of chains of form $s \to i \to t$, where $s$ and $t$ are the source and sink banks and $i$ is some other bank.

2. *No Intermediation:* No bank both holds and writes a CDS on the same reference entity. The liability graphs for individual reference entities are therefore disjoint unions of (in- or out-) star graphs.

3. *No Counterparty Risk:* For each contract, either the holder or the writer is a highly capitalized bank, i.e., its external assets are significantly (by factor $2 \geq 1 + \varepsilon$, for any relevant $\varepsilon$) higher than its maximum liabilities and thus, they cannot default. Further, only highly capitalized banks are writers of CDSs.

Property 1 and 2 are in stark contrast to much of the prior work on financial networks, which has often *only* considered the liability graph, where either reference entities were ignored altogether or they were treated as mere edge labels, but were not identified as nodes in the network. See our literature review in Section 1. No such approach would be able to capture the computational complexity we illustrate in this paper because the liability graph of our hard instances is always trivial.

Property 3 help us discern the "origin of the complexity" from an economic point of view. The holder of a CDS depends on two banks: the reference entity (this is called *fundamental risk*) and the writer of the contract (this is called *counterparty risk*; see D'Errico et al. (2018)). By property 3 however, unless the holder of the CDS is highly capitalized, the recovery rate of the writer is fixed to 1 (up to $\varepsilon$ errors) so that counterparty risk is only the risk of $\varepsilon$ errors. Thus, counterparty risk does not significantly affect recovery rates.[23] The statement also holds for debt contracts. From this, it follows that computational complexity persists if we neglect counterparty risk and must therefore be driven by fundamental risk:

**Proposition 2.** *All our complexity results (Theorems 1–4) still hold in a variant of the clearing model where the assets of a bank $i$ are defined as*

$$a_i(r) := e_i + \sum_{j \in N} c_{j,i}^{\emptyset} + \sum_{j,k \in N} c_{j,i}^{k}(1 - r_k).$$

The above modified model corresponds to a world where a governmental agency like a central bank guarantees the payment in each and every contract while banks are still in default if they cannot pay their obligations. A model where this is the case for CDSs, but not debt contracts, and where our problems are still hard, was

---

studied by Leduc, Poledna and Thurner (2017).[24] The proof of the above proposition is by revisiting our gadget proofs and is omitted. The proofs become slightly easier in the model without counterparty risk because we do not have to deal with $\varepsilon$ errors at the source bank any more.

Overall, we have seen now that the computational complexity illustrated in this paper is not driven by counterparty risk, but by fundamental risk in CDSs. By the second part of property 3, it is further driven by fundamental risk on the *asset side* rather than the liability side of banks' balance sheets. Mathematically, it does not arise from non-linearity and must therefore arise from non-monotonicity (see 2.3; in the above model without counterparty risk and with all relevant liabilities 1, the update function $F$ is piecewise linear and weakly decreasing in the point-wise ordering). To eventually receive a polynomial-time algorithm and thus bound the complexity from above, it therefore seems promising to study structural restrictions under which monotonicity is restored. This is what we do in the following.

## 6.2   Naked CDSs

Non-monotonicity of the update function emerges because a bank that holds a CDS and no other contracts profits from an ill-being of the reference entity. Economically, we say that it is *short* on the reference entity. This effect is only present when CDSs are held by banks in a *naked* fashion, i.e., without holding a corresponding debt contract from the reference entity. The opposite is called a *covered* CDS. In general networks, we need to consider all potential CDS writers to define what a covered and a naked CDS position are. We thus arrive at the following technical definition from our prior work.

**Definition 8** (Covered and Naked CDS Position; Schuldenzucker, Seuken and Battiston (2019))**.** Let $X = (N, e, c, \alpha, \beta)$ be a financial system. A bank $j$ has a *covered CDS position* towards another bank $k$ if

$$\sum_{i \in N} c_{i,j}^k \leq c_{k,j}^{\emptyset}.$$

Otherwise, $j$ has a *naked CDS position* towards $k$. $X$ *has no naked CDSs* if no bank has a naked CDS position towards another bank.

If $j$ has a covered CDS position towards $k$ and the recovery rate of $k$ decreases, then $j$ may receive a higher payment in the CDSs it holds on $k$ (this depends on the recovery rates of the CDS writers), but it also receives a lower payment in the debt contract *from k* and the latter effect weakly dominates. Hence, $j$ can never profit from the ill-being of $k$. A covered CDS thus functions as an insurance against default, while a naked CDS is often considered speculation on default.[25]

For an example for a financial system without naked CDSs, see Appendix A. This also shows that it can still be the case that all solutions are irrational even without naked CDSs, so we still need to consider an approximation problem still.

We have shown in our prior work (Schuldenzucker, Seuken and Battiston, 2019) that in a financial system without naked CDSs, the update function is point-wise monotonically increasing and that this implies that a solution always exists and a simple iteration sequence converges to a solution. It is easy to see that it does so in polynomial time:

**Theorem 6.** *For any financial system $X = (N, e, c, \alpha, \beta)$ without naked CDSs and for any $\varepsilon > 0$, the iteration sequence $(r^n)$ defined by $r^0 = (1, \dots, 1)$ and $r^{n+1} = F(r^n)$ reaches an $\varepsilon$-approximate fixed point of the update function $F$ after $|N| \cdot 1/\varepsilon$ steps. In particular, this defines is a fully polynomial-time approximation scheme (FPTAS) for the total search problem of finding an $\varepsilon$-solution in a financial system with no naked CDSs.*

*Proof.* In each step where $r^n$ is not an $\varepsilon$-approximate fixed point of $F$, some component $r_i^n$ must decrease by at least $\varepsilon$ in the next step. This follows from the definition of an $\varepsilon$-approximate fixed point and monotonicity of $F$. Since $r$ is bounded below by $(0, \dots, 0)$, there can be at most $|N| \cdot 1/\varepsilon$ such steps. This defines an FPTAS because evaluating $F$ and testing for an $\varepsilon$-approximate fixed point can obviously be done in polynomial time and any $\varepsilon$-approximate fixed point of $F$ is an $\varepsilon$-solution (Proposition 1). □

The above result extends to a slightly larger class of networks. In Schuldenzucker, Seuken and Battiston (2019), we have defined a structure called the *colored dependency graph* of a financial network. The nodes of this graph are the banks and an edge $i \to j$ exists whenever $F_j(r)$ depends on $r_i$ (some of the edges may be false positives). Naked CDS positions are colored red and all other edges are colored green. We have shown in our prior work that, if no cycle in this graph contains a red edge, then a solution is still guaranteed to exist and we receive an approximation algorithm, essentially by iterating $F$ on each strongly connected component in topological order. Theorem 6 implies that this algorithm is an FPTAS for the no-red-containing-cycle case.

---

[25]A covered CDS always acts as insurance, but a naked CDS need not be speculative per se. For example, a bank may hold a naked CDS on an entity that has very strong ties with one of its debtors, so that the CDS holder would still never profit from a default of the reference entity. It might also act as a mere intermediary. Detecting and appropriately handling these "indirectly covered" CDSs is a promising topic for future work, but beyond the scope of this paper.

# 7 Conclusion

In this paper, we have studied the clearing problem in financial networks that consist of debt and credit default swap (CDS) contracts. While in the debt-only case, the clearing problem can be solved in polynomial time, we have shown in this paper that the situation is markedly different if CDSs are allowed. Deciding if an (approximate) solution exists is NP-complete and finding an approximate solution when existence is guaranteed is PPAD-complete. In fact, already determining if a specific bank defaults or finding a consistent set of defaulting banks are hard problems. Hardness is preserved under various structural restrictions, but the case where no *naked* CDSs are present allows an FPTAS.

We can now answer our original question: Are financial networks with debt and CDSs "more complex" than those with only debt? Operationalizing informal "complexity" as computational complexity of the clearing problem, we can conclude: Yes, they are more complex, and in a precisely defined way so: understanding the interactions between banks in financial systems with CDSs is at least as challenging as understanding the structure of Boolean and generalized circuits. The complexity prevents us from even knowing which banks default following a shock. Complexity does not arise due to counterparty risk, but due to fundamental risk on the asset side of banks' balance sheets. If anything like a structural "origin" of the complexity can be called out, it should be naked CDSs positions that occur as part of a cycle of dependencies.

These insights are relevant for regulatory policy. The post-2008 regulatory reforms related to the CDS market predominantly target counterparty risk. For example, *margin requirements* mandate counterparties to keep a "buffer account" from which fluctuations in the contract value are offset. Mandatory use of *central counterparties* (CCPs) re-routes all contracts via a highly capitalized central node. *Portfolio compression* eliminates cycles of liabilities for each individual reference entity.[26] All of these policies aim to reduce counterparty risk, but they do not affect fundamental risk. CCPs and portfolio compression modify the network structure, but they leave all reference entity–holder relationships of non-intermediaries as they are. Our results from Section 6.1 imply that this does not eliminate the kind of complexity we reveal in this paper.

Another policy that will likely not affect the hardness of our problems are regulatory *capital constraints*. In our model, this would mean to require a minimum level $\gamma \in [0, 1)$ of external assets relative to maximum liabilities. Banks then have possible recovery rates in $[\alpha\gamma, 1]$ rather than $[0, 1]$. We believe that it will be straightforward to modify our constructions to re-map the latter to the former interval. This is why capital

---

[26]See Financial Stability Board (2017) for details on the different market reforms. Benos, Wetherilt and Zikes (2013) provide an accessible introduction.

constraints do likely not eliminate the complexity we describe.

What *would* eliminate the complexity, by our results from Section 6.2, is banning all naked CDSs. This idea has been part of the public debate following the 2008 crisis (see, for instance, Soros (2009) and Reuters (2009)). During the European sovereign debt crisis in 2011, such a ban was in fact implemented for the subset of CDSs written on sovereign states. The ban is in effect until this day (European Commission, 2011; European Securities and Markets Authority, 2017).

The policy implications we describe here echo earlier results regarding *existence* of a solution in Schuldenzucker, Seuken and Battiston (2019).

Since the structure of our hard instances is so simple, our results are robust to changes to the details of the model. For example, our model abstracts over special provisions in bankruptcy code that essentially give derivatives priority over other contract types (debt in our model) in case of bankruptcy.[27] As our constructions are not affected by counterparty risk and, in fact, relevant banks only ever write a single contract, priority is not relevant and our results persist. Our results *do* crucially depend on the assumption that all contracts are cleared at the same time. That is why they likely do not transfer to any variant of the *dynamic clearing* model in Banerjee, Bernstein and Feinstein (2018) or to a multi-maturity model (Kusnetsov and Veraart, 2019) when debt and CDSs mature at different points in time.

Future work should study which empirical properties of financial networks may make the clearing problem with CDSs feasible. For example, if the number of reference entities is small compared to the number of banks, we might be able to exploit the fact that with recovery rates of reference entities fixed, the update function is linear and monotonic. A similar approach may be feasible when the share of naked CDSs is positive, but small. All of these properties are incompatible with the constructions in our hardness proofs, which leaves hope that efficient algorithms might be available.

Another important topic for future work are practical algorithms that may not have polynomial worst-case running time, but are fast in practice for realistic problem sizes. Prior work has shown that nonlinear optimization solvers can be effective tools in this regard (Rezakhani, 2018). Alternatively, it might be possible to find fast *combinatorial* algorithms. For example, we might perform iteration over default sets in a similar way to how the well-known Lemke-Howson algorithm performs iteration over supports of strategies to compute a Nash equilibrium. By our results from Section 5, any such algorithm would likely imply a combinatorial algorithm for generalized circuits that iterates over discrete assignments.

---

[27] For details see, for example, Bolton and Oehmke (2015).

**Figure 11** Financial System where the unique solution is irrational. Let $\alpha = \beta = 1$ (no default costs).

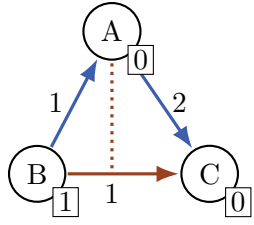

# A  Example That Financial Systems with CDSs May Have Only Irrational Exact Solutions

Figure 11 shows a financial system the unique exact solution of which is irrational. To see this, note that by the contract structure, $a_i(r) \leq l_i(r) \; \forall r, i = A, B$ and therefore $r$ is clearing iff

$$r_A = \frac{r_B}{2}, \qquad r_B = \frac{1}{2 - r_A},$$

and $r_C = 1$. One easily verifies that the unique solution in $[0, 1]^2$ to this system of equations is given by

$$r_A = 1 - \frac{1}{\sqrt{2}}, \qquad r_B = 2 - \sqrt{2}.$$

# B  Comparison of our Generalized Circuit Definition to Rubinstein (2018)

Rubinstein's generalized circuits contain additional gates compared to ours. First, there is a $C_=$ gate that simply copies its input and can of course be replaced by a $C_{\times 1}$ gate. Second, there are additional *Boolean* gates that operate on approximate Boolean values.[28] While we could represent Boolean operations in a financial system using the gadgets from Section 3, Schuldenzucker and Seuken (2019) have shown in prior work that the Boolean operations are in fact redundant and can be represented using the comparison and arithmetic gates. To simplify our analysis, we omit these gates.

The third difference to Rubinstein (2018) is that Rubinstein assumed a binary comparison gate with *two* inputs where $x[v] = 0 \pm \varepsilon$ if $x[a] < x[b] - \varepsilon$ and $x[v] = 1 \pm \varepsilon$ if $x[a] > x[b] + \varepsilon$. One can emulate a binary comparison gate using our unary variant such that $\varepsilon$ increases only by a constant factor. To see this, construct a sub-circuit

---

[28]The definition of approximate Boolean values was weaker than what we did in Section 3, though. See Schuldenzucker and Seuken (2019) for a discussion.

corresponding to the expression

$$\left(\frac{1}{2} + (a - b)\right) - (b - a)$$

and call the output node $u$. Note that the order of operations matters due to truncation at 0 and 1. Then add a $C_{>1/2}$ gate with input $u$ and output $v$. It follows immediately from the gates that if $x$ is an $\varepsilon$-solution, then $x[u] = \tilde{u} \pm 5\varepsilon$ where

$$\tilde{u} := \left[\left[\frac{1}{2} + [x[a] - x[b]]\right] - [x[b] - x[a]]\right] = \left[\frac{1}{2} + x[a] - x[b]\right].$$

Note that $\tilde{u} - 1/2 = \min(1/2, \max(-1/2, x[a] - x[b]))$. From this, it follows that for $\varepsilon \ll 1$ ($\varepsilon < 1/10$ to be precise), $v$ satisfies the definition of the binary comparison gadget for $\varepsilon' := 5\varepsilon$.

## Acknowledgments

We would like thank (in alphabetical order) Vitor Bosshard, Yu Cheng, Constantinos Daskalakis, Timo Mennle, Noam Nisan, and Joseph Stiglitz for helpful comments on this work. Furthermore, we are grateful for the feedback we received from various participants at EC 2016 and ITCS 2017.

All authors gratefully acknowledge financial support from the European Union's FP7 and Horizon 2020 research and innovation programme under Future and Emerging Technologies grant agreements No 610704 (SIMPOL) and No 640772 (DOLFINS). Additionally, Stefano Battiston acknowledges funding from the Swiss National Fund Professorship grant No PP00P1-144689 and from the Institute of New Economic Thinking through the Task Force in Macroeconomic Efficiency and Stability.